\documentclass[12pt,aps,amsmath,latexsym,amsfonts]{JHEP3}
\usepackage[dvips]{epsfig}
\usepackage{epsfig}
\usepackage[all]{xy}
\usepackage{graphicx}

\preprint{LPT-06-63}
\title{
Rotating spacetimes with a cosmological constant}
\author{Christos Charmousis$^{\it 1}$, David Langlois$^{\it{2,3}}$,
Dani\`ele Steer$^{\it{2,1}}$
and Robin Zegers$^{\it {4}}$\\
{\it 1)} LPT, Universit\'e de Paris-Sud, B\^at. 210, 91405 Orsay
CEDEX, France.\\
{\it 2)} Laboratoire APC, B\^atiment Condorcet,
10, rue Alice Domon et Lonie Duquet,
75205 Paris Cedex 13.\\
{\it 3)} Institut d'Astrophysique de Paris, GR$\varepsilon$CO, CNRS, 98bis boulevard Arago, 75014 Paris, France.\\
{\it 4)} Department of Mathematical Sciences,
Durham University,
Science Laboratories,
South Rd.,
Durham,
DH1 3LE.} 
\date{\today}
\abstract{ We develop solution-generating techniques for stationary
metrics with one angular momentum and axial symmetry, in the
presence of a cosmological constant and in arbitrary spacetime
dimension. In parallel we study the related lower dimensional
Einstein-Maxwell-dilaton static spacetimes with a Liouville
potential. For vanishing cosmological constant, we show that the
field equations in more than four dimensions decouple into a four
dimensional Papapetrou system and a Weyl system. We also show that
given any four dimensional ``seed'' solution, one can construct an
infinity of higher dimensional solutions parametrised by the Weyl
potentials, associated to the extra dimensions. When the
cosmological constant is non-zero, we discuss the symmetries of the
field equations, and then extend the well known works of Papapetrou
and Ernst (concerning the complex Ernst equation) in
four-dimensional general relativity, to arbitrary dimensions. In
particular, we demonstrate that the Papapetrou hypothesis
generically reduces a stationary system to a static one even in the
presence of a cosmological constant.  We also give a particular
class of solutions which are deformations of the (planar) adS
soliton and the (planar) adS  black hole. We give example solutions
of these techniques and determine the four-dimensional seed
solutions of the 5 dimensional black ring and the Myers-Perry black
hole.}
\maketitle
\preprint{gr-qc/0610091\\
LPT-ORSAY/0663}

\newcommand{\be}{\begin{equation}}
\newcommand{\ee}{\end{equation}}
\newcommand{\beq}{\begin{equation}}
\newcommand{\eeq}{\end{equation}}
\newcommand{\ba}{\begin{eqnarray}}
\newcommand{\ea}{\end{eqnarray}}
\newcommand{\bea}{\begin{eqnarray}}
\newcommand{\eea}{\end{eqnarray}}
\newcommand{\bean}{\begin{eqnarray*}}
\newcommand{\eean}{\end{eqnarray*}}
\newcommand{\bml}{\begin{mathletters}}
\newcommand{\eml}{\end{mathletters}}
\newcommand{\del}{\overrightarrow{\nabla}}

\def\half{\textstyle{1\over2}}

\def\M{{\cal{M}}}

\def\R{{\cal{R}}}
\def\E{{\cal{E}}}

\def\r{\mathrm{r}}
\def\z{\mathrm{z}}

\def\R{\mathrm{R}}
\def\Z{\mathrm{Z}}
\def\a{a}
\def\U{\hat U}
\def\A{\cal{A}}
\def\q{q}
\def\n{\mathrm{n}}

\newcommand{\e}{{\rm e}}

\newcommand{\lb}{\label}
\newcommand{\bb}{\bibitem}

\begin{document}
\section{Introduction}

Exact solutions in General Relativity are essential in order to gain
insight on the nature of gravity, and for this reason much effort
has been devoted to their systematic construction. In
four-dimensional Einstein general relativity, numerous methods have
been developed to obtain solutions, usually by assuming some
symmetries for spacetime beforehand \cite{SKMHH}. An important class
of such solutions are spacetimes in vacuum which are axially
symmetric and either static or stationary. In the former case, Weyl
\cite{Weyl} showed that spacetime metrics can be generated from
solutions of the Laplace equation in three-dimensional cylindrical
coordinates, and hence that the field equations are essentially
integrable{\footnote{By essential here we
    mean that any solution can be expanded as an infinite series over a
    self-adjoint basis of orthonormal functions; see for example \cite{quevedo}.}}.
Many solutions of physical interest belong to this class: Rindler
spacetime, the Schwarzschild black hole, as well as the C-metric
\cite{cmetric} describing in part  an accelerating black hole, and
multiple black hole solutions \cite{israelkahn}.

The work of Weyl was extended to stationary and axisymmetric
spacetimes by Lewis \cite{lewis} and Papapetrou \cite{Papa1,Papa2}.
Typical examples of such spacetimes are the rotating black hole
solution found in the 60's by Kerr \cite{ker}, and the Taub-NUT (TN)
solution \cite{TN} which has a new charge and non-trivial spacetime
asymptotical behaviour. A great deal of work has also been devoted
to developing and extending solution generating methods, and then to
the analysis of the resulting new solutions: see \cite{SKMHH} for a
review of this vast subject and references therein.

In this paper we focus on the powerful methods first developed by
Ernst \cite{Ernst}.  Their extension enabled relativists (see
\cite{SKMHH}, \cite{bz}, \cite{piotr} and references within) to
demonstrate that, for vanishing cosmological constant, stationary
and axisymmetric metrics are also essentially integrable. Although
there have  been an important number of papers on the subject,
little is known when one includes the cosmological constant in
Einstein's field equations.  As we shall discuss in detail, the
system is no longer integrable in this case{\footnote{This is true
even for
    Weyl's static case.}},
and methods such as those introduced by Papapetrou and Ernst, at
first glance, seem to fail. In rather simple terms, integrability
breaks down because equations which were homogeneous for $\Lambda=0$
become inhomogeneous when $\Lambda \neq 0$. Examples of interesting
stationary axisymmetric solutions with $\Lambda \neq 0$ are scarce:
Carter, for example, found the extension of Kerr's solution with a
cosmological constant (as well as a Taub-NUT parameter) by
considering separable ans\"atze for Einstein's equations
\cite{carter}.

With the advent of modern theories of unification and in particular
string theory, interest in solutions and solution generating methods
in higher dimensional gravity has gradually developed. Myers and
Perry \cite{MP} first gave the extension of Kerr's solution to
higher dimensions, whereas extensions of Carter's solution were
undertaken in \cite{hawking}. In parallel, given the $p$-brane
solutions of Horowitz and Strominger \cite{HS} and their importance
in the understanding of string theory \cite{Polchinski}, much work
has been devoted to Einstein-Maxwell-Dilaton (EMD) theories.  When
an EMD solution is Weyl symmetric (i.e.~static and axisymmetric) it
can, via an exact Kaluza-Klein mechanism and for certain values of
the coupling constants appearing in the action, be uplifted to a
higher dimensional axisymmetric and stationary vacuum solution. An
example is the 4 dimensional Reissner-Nordstrom solution which can
be mapped to a 5 dimensional rotating black 1-brane, and for which
the black hole charge turns into the rotation potential and
vice-versa. Rather less trivially, the work of Dowker et
al.~\cite{D93} in four dimensions, where the C-metric was upgraded
to an EMD solution, allowed Emparan and Reall to discover the black
ring solution{\footnote{See \cite{ER3} for a full review on black
ring type of  solutions and a full list of references on the
subject.}} in 5 dimensions \cite{ER} (see also \cite{ERS} for a
supersymmetric version). This solution represents a rotating black
hole of given mass and angular-momentum, with a horizon of ring
topology, $S^2 \times S^1$, thus making it different to the
Myers-Perry solution. Indeed, the black ring is a typical higher
dimensional solution preventing the extension of 4 dimensional
uniqueness theorems \cite{carter2}{\footnote{However, see
\cite{wald2}, where it is shown that a stationary rotating black
hole must have an axial Killing vector.}}. Regarding solution
generating methods (see \cite{Coley} for work on classification of higher dimensional solutions), Emparan and Reall \cite{ER02} extended Weyl's
work to higher dimensions while a cosmological constant was included
in the analysis of \cite{Charmousis:2003wm}. Recently Harmark et
al.~\cite{Harmark,Olesen} analysed stationary and axisymmetric
metrics for $\Lambda=0$, giving the relevant mappings of solutions
in multiple coordinate systems.

This paper aims to study solution generating methods for
stationary and axisymmetric spacetimes in arbitrary dimension, and
with non-vanishing cosmological constant $\Lambda \neq 0$. Apart
from the interest in classical and higher dimensional general
relativity, one must stress the importance of asymptotically adS
solutions in string theory. Any such solution is a classical
background with which to put the adS/CFT correspondence to the
test \cite{maldacena}. Furthermore, recent exotic developments in
cosmology, such as braneworlds, have brought particular attention
to gravitating solutions of axial symmetry in adS.
 Indeed, an axially symmetric metric in 5 dimensions
corresponds to a spherically symmetric geometry on the brane. A
solution that  describes a 4 dimensional black hole localised on a
Randall-Sundrum (RS) braneworld \cite{RS} (see \cite{EHM} for a
clear explanation in lower dimensions, and also \cite{ruth} and
references within), if it exists would enter this category. One in
particular, would seek a very particular metric of axial symmetry:
the equivalent of a C-metric in 4 dimensions which describes, in
part, an accelerating black hole. The reason for this is the
following: an RS brane, embedded in a negatively curved spacetime,
is charted in Poincar\'e coordinates, so that the brane induced
metric is flat. This coordinate system from the bulk point of view
is an accelerating patch covering a part of adS space. In rather
loose terms this patch is similar for adS to the Rindler coordinates
for Minkowski spacetime. Therefore a localised RS black hole has to
be accelerating in order to keep up with the brane, meaning in turn
that in the 5 dimensional bulk one wants a generalised C-metric:
such a solution is yet unknown, even when $\Lambda=0$ (see
\cite{Charmousis:2003wm} for a recent discussion). More generally,
axisymmetric solutions are important for theoretical, related in
particular to the issue of stability of higher dimensional black
hole solutions \cite{GL}, and phenomenological reasons, particularly
in the context of detecting extra dimensions. Furthermore, they are
also related to solutions describing anisotropic Bianchi type
cosmologies with perfect fluid sources (see \cite{sphere}) or again
to the gravitational field of sources such as the linear cosmic
string \cite{davis}.  It was found in \cite{davis} that the weak
field approximation for the metric around a RS localised cosmic
string differs from the 4 dimensional one \cite{vilenkin}. These
questions are even more intriguing since recent work \cite{kal},
making use of the adS/CFT correspondance relating such bulk
backgrounds with their brane-boundaries, can promote such classical
solutions as probes of quantum effects on the braneworld.

In this paper we consider $D$-dimensional Einstein gravity with a
cosmological constant term, and search for stationary and
axisymmetric solutions. From a Kaluza-Klein perpsective we also
 consider $d=D-1$ dimensional EMD solutions (see, for example,
\cite{cedric}) with a Liouville potential for the dilaton (see, for
example, \cite{christos}).  For simplicity we consider a single
angular momentum parameter throughout, thus postulating the
existence of $D-2$ Killing vectors of which only two are non
orthogonal. We begin in section \ref{sec:over} by reviewing
stationary and axisymmetric spacetimes with $\Lambda=0$ in 4
dimensions, and also the well known solution generating methods of
 Ernst and Papapetrou. Then, in section \ref{sect:Dim}, we
introduce the cosmological constant and generalise the
dimensionality of spacetime.  The field equations are set up in a
convenient form which resembles (but is not identical to) the
original  Lewis-Papapetrou 4 dimensional form, and this enables us
to discuss their symmetries and extend 4 dimensional electromagnetic
duality to include the presence of a cosmological constant.
Furthermore it allows us to generalise the Ernst equation to
arbitrary $d$ and $\Lambda \neq 0$ (subsection \ref{subsec:ernst});
to extend Papapetrou's method (for arbitrary $d$ and $\Lambda \neq
0$ )and demonstrate that any Weyl solution  gives a class of
rotating solutions satisfying Papapetrou's hypothesis (subsection
\ref{subsec:papa}); to give a special class of solutions which
describe deformations of the adS soliton and the planar adS black
hole; and finally to present a method which allows for the direct
construction of higher dimensional rotating metrics from lower
dimensional ones (subsection \ref{subsec:Dcon}). Finally, in
sections \ref{sec:4}, \ref{sec:5} and \ref{sec:6}, we give some
simple examples and put into practice the methods developed.
Conclusions are given in section \ref{sec:conc}.

\section{An overview of $D=4$ axially symmetric solutions of the
vacuum}\label{sec:over}

In four dimensions, a static and axisymmetric metric can be written in the form
\beq
ds^2=-e^{2\lambda}dt^2+e^{-2\lambda}\left[\alpha^2 d\varphi^2+e^{2\chi}(d\R^2+d\Z^2)\right],
\label{W0}
\eeq
where $\alpha$, $\lambda$ and $\chi$ are functions of $\R$ and $\Z$
only.  It follows from the vacuum Einstein equations $R_{ab}=0$
that $\alpha$ is harmonic,
$\Delta \alpha = (\partial_\Z^2 + \partial_\R^2) \alpha =0$, and hence
that one can always set $\alpha=\r$ by a two dimensional conformal
transformation in the $(\R,\Z)$ plane. Without loss of generality,
the metric then takes the well-known Weyl form \cite{Weyl}
\beq
ds^2=-e^{2\lambda}dt^2+e^{-2\lambda}\left[\r^2d\varphi^2+e^{2\chi}(d\r^2+d\z^2)\right],
\label{WC}
\eeq
and in this special coordinate system
$\lambda(\r,\z)$ now satisfies
\beq
\label{weylrod} \left(
\partial_{\r}^2 +\frac{1}{\r}\partial_{\r} +\partial_{\z}^2\right) \lambda=0.
\eeq Since this is just the three-dimensional flat Laplace equation
in cylindrical coordinates, formally $\lambda$ can be seen as the
Newtonian potential generated by an axisymmetric Newtonian source
\cite{Weyl}. Once a solution (or potential) $\lambda$ is chosen in
(\ref{weylrod}), the full metric is determined by solving the
remaining Einstein's equations for $\chi$:
 \beq \label{nonlinear}
\partial_{\r}\chi=\r\left[\left(\partial_{\r}\lambda\right)^2-
\left(\partial_{\z}\lambda\right)^2\right], \qquad
\partial_{\z}\chi=2\, \r\, \partial_{\r}\lambda\partial_{\z}\lambda,
\eeq
which carry the full non-linearity of $R_{ab}=0$. Since
(\ref{weylrod}) is linear,
 one can
superpose $\lambda$-potentials and then calculate the relevant
$\chi$ field from (\ref{nonlinear}).  For instance, the
Schwarzschild solution corresponds to the Newtonian potential of a
rod placed at $z=0$ and of finite length (per unit mass) in the $z$
direction (see for example \cite{israelkahn} or
\cite{Charmousis:2003wm}); the Rindler spacetime corresponds to a
semi-infinite rod; and their superposition gives rise to the
Newtonian potential corresponding to the C-metric describing, in
part, the spacetime of an accelerating black hole \cite{cmetric}.
This is one intuitive way of obtaining solutions in the form
(\ref{WC}). Alternatively it is useful to recall that since
(\ref{weylrod}) is a linear second order equation one can solve it
directly by separation of variables, find the relevant
eigenfunctions for the separate Sturm-Liouville problems, and then
expand in terms of the basis of functions (see \cite{quevedo}).

The choice of a coordinate system in which to undertake the task
of writing down the metric solutions can be crucial. Although the
Weyl canonical form is particularly helpful for the analysis of
the system of equations at hand and for classifying the solutions,
 it is often  useful to write specific solutions in coordinates
differing from  those in (\ref{WC}). A particularly appropriate
coordinate system turns out to be the spheroidal coordinates
discussed by Zipoy \cite{Zipoy}, which have ellipsoids and
hyperboloids of revolution as coordinate surfaces. As we will see
below, they are tailored to describe the Schwarzschild Weyl
potential
 and were first introduced in order to express the
exact Newtonian potential around the earth.
Thus rather than Weyl coordinates $(\r,\z)$, consider polar-like
coordinates $(u,\psi)$ but with hyperbolae as radial functions,
that is
\bea
\z&=& \cosh u\, \cos\psi\, , \nonumber\\
\r&=& \sinh u \, \sin\psi\, , \label{prolate} \eea so that in the
$(\r,\z)$ plane $\psi={\rm const}$ curves are hyperboloids and
$u={\rm const}$ are ellipsoids. On setting $x=\cosh u$ and
$y=\cos\psi$, the coordinate system becomes symmetric in $x$ and
$y$: the 2 dimensional line element is given by \beq d\r^2+d\z^2 =
(x^2-y^2)\left[\frac{dx^2}{x^2-1} + \frac{dy^2}{1-y^2} \right]
\label{line} \eeq and the  Laplace equation (\ref{weylrod}) takes
the form \beq \frac{1}{x^2-y^2} \left\{ \frac{\partial}{\partial x}
\left[(x^2-1) \frac{\partial \lambda}{\partial x} \right] +
\frac{\partial}{\partial y} \left[(1-y^2) \frac{\partial
\lambda}{\partial y} \right] \right\}=0. \label{sp} \eeq As an
example of these different coordinate shuffles and one in which
spheroidal coordinates appear naturally, consider a Schwarzschild
black hole: the standard metric
\be
ds^2=-\left(1-{2M \over r}\right)dt^2+\frac{dr^2}{1-{2M\over r}}+r^2 d\Omega_{II}^2
\ee
can be rewritten in Weyl coordinates $(\R,\Z)$ of (\ref{W0}) where
$r/2M=\cosh^2(\R/2)$ and $\theta=\Z$. The conformal transformation
to (\ref{WC}) then gives $\z=\cos\Z \cosh \R$ and $\r = \sin \Z
\sinh \R$ as in (\ref{prolate}) (that is, $u=\R$ and $\psi=\Z$), and
\be \label{5} e^{2\lambda}=\frac{x-1}{x+1}. \ee It can be easily
checked that this Weyl potential $\lambda$ is indeed a solution of
(\ref{sp}).  More generally, the solutions of (\ref{sp}) are
separable and consist of products of Legendre polynomials
\cite{Zipoy}. Appropriate boundary conditions (as well as other
coordinate systems) have been considered by different authors
\cite{quevedo,Zipoy}. Spheroidal coordinates are also very relevant
for the analysis of {\it stationary} axisymmetric vacuum solutions,
as we now discuss.

Lewis and Papapetrou  generalised the  approach of Weyl to
stationary and axisymmetric solutions in vacuum
\cite{lewis,Papa1,Papa2}. After a conformal transformation, the
metric takes the Lewis-Papapetrou form \beq
ds^2=-e^{2\lambda}\left(dt+Ad\varphi\right)^2+e^{-2\lambda}
\left[\r^2d\varphi^2+e^{2\chi}(d\r^2+d\z^2)\right], \label{stat}
\eeq which differs from the static form by  the additional component
$A=A(\r,\z)$. Note that $\partial_t$ is no longer a static but
rather a stationary (locally) timelike Killing vector field, and
that one cannot, via a coordinate transformation, remove the
non-diagonal metric component whilst keeping the line-element `$t$'
independent. For the metric (\ref{stat}), Ernst \cite{Ernst} pointed
out an interesting reformulation of  Einstein's equations for $A$
and $\lambda$, which read respectively \beq \label{ZZ}
\partial_{\r}\left(\frac{e^{4\lambda}}{\r}\partial_{\r}A\right)+\partial_{\z}\left(\frac{e^{4\lambda}}{\r}\partial_{\z }A\right)=0,
\qquad \left(
\partial_{\r}^2+\frac{1}{\r}
\partial_{\r}+\partial_{\z}^2\right) \lambda=\frac{e^{4\lambda}}{2\, \r^2}\left[(\partial_{\r}A)^2+(\partial_{\z}A)^2
\right]. \nonumber
\eeq
Indeed, on introducing an auxiliary field, $\omega$,
defined{\footnote{As we will see
    later on, this auxiliary field describes nothing but the passage from an
    electric to a magnetic potential and vice-versa.}} by
\beq
\label{zoo}
\left(-\partial_{\z}\omega, \partial_{\r}\omega\right)=\frac{e^{4\lambda}}{\r}
\left(\partial_{\r} A,\partial_{\z} A\right),
\eeq
the complex function
\beq
\E=e^{2\lambda}+i\omega \label{TPs}
\eeq
then satisfies the complex differential equation
\beq
\frac{1}{\r} \del \cdot \left (\r \del \E \right ) = \frac{(\del
  \E)^2}{\mbox{Re}(\E)}
\label{E4}
\eeq
known as the Ernst equation. Its real and imaginary part are exactly
(\ref{ZZ}).\footnote{ Note that here, $\del=(\partial_\r,
\partial_\z)$, whereas in the literature $\del$ is often used to
denote a three-dimensional gradient in cylindrical coordinates.}
In this language, the Weyl potential $\lambda$ is simply given by
the real part of the Ernst potential $\E$, whereas rotation is
embodied  by a non-trivial $\omega$.

Using the symmetries of complex functions, several methods have
been proposed to obtain solutions of the Ernst equation (\ref{E4})
and hence to generate new metrics (see \cite{SKMHH}, \cite{Ernst},
\cite{piotr} and references within). An elegant application
appeared in Ernst's original paper \cite{Ernst}, namely a simple
method to obtain the Kerr solution from the Schwarzschild
solution. This example  also underlines the importance of the
choice of coordinates. Indeed, let
\be
\E=\frac{\xi-1}{\xi+1}. \label{xidef}
\ee
Then in spheroidal coordinates and for the Schwarzschild solution,
it follows from (\ref{5}) and (\ref{TPs}) that $\xi=x$. Note that
our new metric component $\xi$ is now the 'radial' coordinate $x$,
rather as in (\ref{WC}) where $\alpha=\r$. We have adapted the
coordinate system to the real part of the black hole Ernst
potential. By symmetry, $\xi=y$ is also solution of (\ref{E4}), as
is $\xi=x \sin\vartheta +i y \cos\vartheta$. It turns out that
this is nothing other than the Ernst potential of the Kerr black
hole, where $\sin\vartheta= a/M$ is the ratio between the angular
momentum parameter and the mass of the black hole \cite{Ernst}.

In a similar manner, Papapetrou noted that if one makes the
hypothesis $\lambda=\lambda(\omega)$  then the system (\ref{ZZ}),
with (\ref{zoo}), is integrable \cite{Papa1}. Solutions obtained
this way generally have non-trivial asymptotic properties and, in
particular, Gautreau and Hoffman \cite{GH} showed that the above
hypothesis reduces the stationary Papapetrou system to a Weyl
static system. They also showed that starting from the Weyl
potential of the Schwarzschild black hole one could easily
construct the TN solution \cite{TN}, which thus belongs to the
Papapetrou class.

We now proceed to generalise the work of Ernst and Papapetrou to higher
dimensions, including a non-vanishing cosmological constant.

\section{Rotating spacetimes and the Einstein-Maxwell-dilatonic
 (EMD) system} \label{sect:Dim}

\subsection{Set-up of the field equations and their symmetries}

We consider $D$-dimensional stationary axisymmetric metrics of the
form \beq ds_D^2=- e^{2W}\left(dt+A\, d\varphi\right)^2+
e^{2U_\varphi}d\varphi^2+ \sum_{i=1}^{D-4} e^{2U_i} dx_i^2+e^{2V}
(d\r^2+d\z^2) \label{metric} \eeq where all the metric components
are functions of $\r$ and $\z$ only, and we search for solutions of
the vacuum Einstein equations with a cosmological constant \beq
G_{AB}+\Lambda g_{AB}=0. \eeq The metric (\ref{metric}) possesses
$(D-2)$ Killing vector fields, of which $\partial_t$ and
$\partial_\varphi$ are not orthogonal to each other, so that the
spacetime is stationary rather than static. When $D=4$,
(\ref{metric}) is the most general stationary axisymmetric metric
(which, when $\Lambda=0$, can be written in the form (\ref{stat})).
For $D>4$ multiple angular momenta are possible: here, however, we
work with (\ref{metric}) which can be seen as the simplest
generalization, through the addition of a single angular momentum
$A$, of a static axisymmetric $D$-dimensional Weyl solution.

For the following analysis, it will be useful to recall (see for example \cite{KKBH}) that
(\ref{metric}) can be dimensionally reduced to a $(D-1)$
dimensional EMD system. Numerous higher-dimensional solutions have been obtained this way, \cite{KKBH}, \cite{Cvetic}. Indeed,
Kaluza-Klein reduction of the metric (\ref{metric}) yields,
putting aside the question of the signature
 for the moment, a $(D-1)$-dimensional
metric together with a scalar field and a vector potential. More
explicitly, if one starts from a $D$-dimensional metric $\tilde
g_{AB}$, with dynamics governed by \beq S_{D}=\int d^Dx\sqrt{-\tilde
g}\left(\tilde R-2\Lambda\right) \, , \label{actionvacuum} \eeq
 and
decompose $\tilde{g}_{AB}$ as
\beq \tilde{ds}^2_D=e^{-2\a
\phi}ds_{D-1}^2+e^{2(D-3)\a\phi}(dw+A_\nu
 dx^\nu)^2 . \label{KKmetric}
\eeq
then the $(D-1)$-dimensional metric $g_{\mu \nu}$, the $(D-1)$
form $A_\nu$ and the scalar field $\phi$ obey the system of
equations derived from the action \beq S_{D-1}=\int d^{D-1}x
\sqrt{-g} \left[R-(D-2)(D-3)\a^2(\partial\phi)^2 -\frac{1}{4}
e^{-2(D-2)\a\phi} F^2-2\Lambda e^{2\a\phi}\right], \label{KKaction}
\eeq with field strength \beq F_{\mu \nu}= \partial_{\mu }A_{\nu} -
\partial_{\nu }A_{\mu} . \eeq Note that the dependence on the
dilaton, $\phi$, in (\ref{KKmetric}) has been chosen so that the
$(D-1)$ dimensional action (\ref{KKaction}) corresponds to
 the Einstein frame. Notice also that since
$\Lambda \neq 0$, the dilaton acquires an exponential potential. We
now set
\beq \a=\pm \frac{1}{\sqrt{2(D-2)(D-3)}} \eeq so that the
kinetic term for $\phi$ is canonically normalised, and in turn the
dilaton's potential and its coupling to the field strength are
completely determined.

We now generalise one step further and, rather than
(\ref{KKaction}), consider the action \beq S_{d}=\int d^{d}x
\sqrt{-g} \left[R-\frac{1}{2}(\partial\phi)^2 -\frac{1}{4}
e^{\gamma\phi} F^2-2\Lambda e^{-\delta\phi}\right], \label{action}
\eeq where
 now the parameters $\gamma$ and $\delta$ are {\it
arbitrary}. Solutions to (\ref{action}) have been studied in the
past (see e.g. \cite{Gib}) including spacetimes with non-trivial
asymptotic behaviour \cite{cedric}. Broadening the parameter space
in this way will enable us to study the generic properties of the
system of equations derived from (\ref{action}), which are the
subject of the remainder of this paper. Indeed, it is worth
stressing that the black ring solution \cite{ER}, which is a
five-dimensional vacuum solution, was derived from a
four-dimensional solution of an Einstein-Maxwell dilatonic system
\cite{D93}.  The solutions of (\ref{action}) are, of course,
solutions of the $D$-dimensional action (\ref{KKaction})
with{\footnote{Throughout this paper we will note by $D$ the
dimension of
  the uplifted metrics, whereas $d$ will denote
  the dimension of the EMD spacetime.}}
\be
d=D-1, \label{ddef}
\ee
if the coupling parameters take the specific values
\beq
\gamma = \pm \sqrt{\frac{2(D-2)}{(D-3)}} \; , \qquad \delta = \pm
\sqrt{\frac{2}{(D-2)(D-3)}}=2a. \label{gd}
\eeq
From (\ref{metric}), the $d$-dimensional metric $g_{\mu \nu}$ in
(\ref{action}) is fully diagonal and a Weyl metric.

We suppose here that the vector potential $A_\mu$ in (\ref{action})
has only one non-zero component since we only consider a single
angular momentum for the uplifted case. This non-zero component
$A_{\mu_*}$ can be timelike  ($\mu_*=0$), in which case  the vector
potential $A_\mu$ is said to be electric, whereas if it is spacelike
($\mu_*\neq 0$), $A_\mu$ is magnetic. In both cases, we consider a
diagonal $d$-dimensional metric of the form \beq
ds^2_d=-e^{2U_0}dx_0^2+\sum_{i=1}^{d-3}e^{2U_i}dx_i^2+e^{2V}(d\r^2+d\z^2),
\label{Weyll1} \eeq where the functions $U_\mu$ (with $\mu=0,\dots,
d-3$) and $V$ only depend on $\r$ and $\z$.

When the $d$-dimensional EMD solution is related to a $D=d+1$
dimensional vacuum solution, i.e.~when the coupling parameters
satisfy (\ref{gd}), then an electric solution can be uplifted to a
$D$-dimensional metric of the form (\ref{metric}) via a double Wick
rotation of the metric (\ref{KKmetric})\ba \label{wiw}
w &\rightarrow & i t \\
x_0 &\rightarrow & i \varphi.
\ea
 In the magnetic case, one must not only use the double Wick rotation
\ba
w &\rightarrow & i t \\
x_0 &\rightarrow & i y \ea where $y$ is one of the space-like
coordinates $x_i$ in (\ref{metric}), but also transform $A_{\mu_*}$
according to \be A_{\mu_*} \rightarrow i A \, . \ee The case $d=3$
is rather special since there is only one extra coordinate other
than $\r$ and $\z$. In the magnetic case, the extra coordinate, say
$x_1$, is necessarily spacelike which implies that the 3-dimensional
metric (\ref{Weyll1}) is {\it a priori} of Riemannian signature.
One can then obtain a Lorentzian $D=4$  metric of the form
(\ref{metric}) via the transformations $w \rightarrow  i t$ and
$A_{\mu_*}=A_1 \rightarrow i A $.

 When $d=4$ and $\Lambda=0$ the
electric and magnetic spacetimes are linked via the well-known
electromagnetic duality relating strong to weak dilaton coupling,
namely \beq \label{EM} \phi\rightarrow \bar\phi=-\phi, \qquad
F_{\mu\nu}\rightarrow \bar F_{\mu\nu}=\frac{1}{2}e^{\gamma \phi}
\epsilon_{\mu\nu\rho\sigma}F^{\rho\sigma} . \eeq We will discuss
duality relations for $\Lambda \neq 0$ at the end of this section.

Given these well-known preliminaries and notation issues, we are now
ready to analyse the equations of motion coming from (\ref{action})
with $g_{\mu \nu}$ given in (\ref{Weyll1}). It is useful to define
\beq \label{sphere} \alpha=\exp \left ( \, \sum_{\mu=0}^{d-3}U_\mu\,
\right ), \quad \U_\mu=U_\mu-\frac{1}{d-2}\ln\alpha, \quad
\chi=V+\frac{d-3}{2(d-2)}\ln\alpha, \eeq so that the deviations,
$\U_\mu$, from the average, $\alpha$, sum to zero: \be
\sum_{\mu=0}^{d-3}\U_\mu=0. \label{dev} \ee In terms of these
functions the metric (\ref{Weyll1}) is given by \be ds^2_d    =
e^{2\chi} \alpha^{-\frac{d-3}{d-2}} \left (d\r^2 + d\z^2 \right ) +
\alpha^{\frac{2}{d-2}} \sum_{\mu=0}^{d-3} \eta_{\mu\mu} e^{2\U_\mu}
(dx^\mu)^2 \, , \ee where $\eta_{\mu\nu}$ is the Minkowski metric.
Let us also introduce the complex conjugate coordinates $u$ and $v$
such that \be u=\frac{{\r}-i {\z}}{2}, \qquad
v=\frac{{\r}+i{\z}}{2},\qquad {\rm and} \qquad 4 \, du
dv=d{\r}^2+d{\z}^2, \qquad {\r},{\z}\in {\mathbb R} . \ee Then the
equations of motion derived from (\ref{action}) are \ba
\label{WeylKK1}
\Delta \alpha &=& -2\Lambda \alpha^{\frac{1}{d-2}} e^{2\chi-\delta \phi}, \\
\label{WeylKK2}
0 &=& \overrightarrow{\nabla} \cdot \left (
e^{\gamma \phi -2\U_*} \alpha^{\frac{d-4}{d-2}} \overrightarrow{\nabla} {A} \right ),  \\
\label{WeylKK3} \frac{1}{\alpha} \overrightarrow{\nabla} \cdot
\left ( \alpha \overrightarrow{\nabla} {\phi} \right ) &=&
\frac{\gamma\epsilon}{2} e^{\gamma \phi} \alpha^{-\frac{2}{d-2}}
e^{-2\U_*} \left (\overrightarrow{\nabla} A \right )^2 -2\delta
\Lambda \alpha^{-\frac{d-3}{d-2}}
e^{2\chi-\delta \phi}, \\
\label{WeylKK4}
\frac{1}{\alpha} \overrightarrow{\nabla} \cdot \left ( \alpha
  \overrightarrow{\nabla} {\U_*} \right ) &=&  -\epsilon \frac{d-3}{2(d-2)}
  e^{\gamma \phi-2\U_*} \alpha^{-\frac{2}{d-2}}\left
(\overrightarrow{\nabla} A \right )^2, \\
\label{WeylKK5}
\frac{1}{\alpha} \overrightarrow{\nabla} \cdot \left ( \alpha
  \overrightarrow{\nabla} {\U_{\hat\mu}} \right ) &=&  \frac{\epsilon}{2(d-2)}
  e^{\gamma \phi-2\U_*} \alpha^{-\frac{2}{d-2}}\left
(\overrightarrow{\nabla} A \right )^2 \qquad , \; ({\hat\mu}\neq \mu_*)\\
\label{WeylKK6}
2 \chi_{,u} \frac{\alpha_u}{\alpha}- \frac{\alpha_{,uu}}{\alpha} &=&  {\bf
  \U}_{,u}^2 + \frac{1}{2} \phi_{,u}^2
+\frac{\epsilon}{2} e^{\gamma \phi} \alpha^{-\frac{2}{d-2}}
e^{-2\U_*} \left ( A_{\, ,u} \right )^2 \qquad (u\leftrightarrow v),
\ea where we have distinguished the component $\U_*\equiv
\U_{\mu_*}$ (along the direction in which the potential $A_\mu$ is
switched on) from the other components denoted by $\U_{\hat\mu}$. An
extra equation exists for $\chi$ but it is just a Bianchi identity
so we omit it. The parameter $\epsilon$ takes the value
$\epsilon=-1$ when the potential is electric, and $\epsilon=1$ when
it is magnetic. Equation (\ref{WeylKK2}) is simply Maxwell's
equation, whilst equation (\ref{WeylKK3}) is the equation of motion
for the dilaton. Finally, the
ordinary (complex)
differential equation (\ref{WeylKK6}) and its complex conjugate,
where we have set ${\bf \U}_{,u}^2 =
\sum_{\mu=0}^{d-3}{U}_{\mu,u}^2$, yield two real partial
differential equations by restriction to their real and imaginary
parts.

For the following analysis it is expedient to rewrite equations
(\ref{WeylKK1})-(\ref{WeylKK6}) in a form as close as possible to
the original Papapetrou and Ernst formulation of the $D=4$ equations
of motion with $\Lambda=0$ (section \ref{sec:over}). To do so we
follow the following strategy: decouple whenever possible the field
equations between them; use (\ref{WeylKK1}) to absorb the
cosmological constant $\Lambda$; and finally render the field
equations as independent of the dimension $d$ as possible. Consider
therefore the linear combinations \ba \label{trans}
\Psi_{\mu_*}\equiv \Psi_{*} &=&
\sqrt{\frac{d-3}{d-2}}\left[\sqrt{\frac{d-3}{d-2}}(\phi-\delta
\ln\alpha)+\gamma\U_*\right],
\qquad   \; (d>3) \\
\Psi_{\hat\mu} &=& \, \U_{\hat\mu}+\frac{1}{d-3}\U_*, \qquad \qquad  \qquad \qquad \; \;\qquad  \qquad  (d>3)\nonumber\\
\Omega &=& \gamma(\phi-\delta \ln\alpha)-2\U_*\nonumber \\
2\nu &=& 2\chi-\delta\phi + \frac{\delta^2}{2}\ln\alpha\nonumber ,
\ea and we take  $\Psi_\mu=0$ for $d=3$. From (\ref{dev}), it
follows that $\sum_{{\hat\mu}\neq \mu_*}\Psi_{\hat\mu}=0$. On
defining the positive constant \be \label{s} s \equiv
\gamma^2+2\frac{d-3}{d-2} \, , \ee the equations
(\ref{WeylKK1}-\ref{WeylKK6}) simplify to \ba \label{WKK1}
\Delta \alpha &=& -2 \Lambda \alpha^{\frac{1}{d-2}-\frac{\delta^2}{2}} e^{2\nu} \, , \\
\label{WKK2}
0 &=& \overrightarrow{\nabla} \cdot \left ( e^{\Omega}
  \alpha^{\frac{d-4}{d-2}+\gamma \delta} \overrightarrow{\nabla} {A} \right ) \, , \\
\label{WKK3} \frac{1}{\alpha} \overrightarrow{\nabla} \cdot \left
( \alpha \overrightarrow{\nabla} {\Omega} \right ) &=&
\frac{\epsilon s}{2} e^{\Omega}
\alpha^{\gamma\delta-\frac{2}{d-2}} \left (\overrightarrow{\nabla} A \right )^2 \, , \\
\label{WKK4}
\overrightarrow{\nabla} \cdot \left ( \alpha
  \overrightarrow{\nabla} {\Psi_\mu} \right ) &=&  0 \, , \qquad (\mu=0, \dots, d-3)\\
\label{WKK5} 2 \nu_{,u} \frac{\alpha_u}{\alpha}-
\frac{\alpha_{,uu}}{\alpha} &=& \frac{1}{s} \left(\Psi_{*,u}^2
+\frac{1}{2}  \Omega_{,u}^2 \right) +\frac{\epsilon}{2} e^{\Omega}
\alpha^{\gamma\delta -\frac{2}{d-2}}  \left ( A_{\, ,u} \right )^2+\nonumber\\
&+& \; \sum_{{\hat\mu}\neq \mu_*}\Psi_{{\hat\mu},u}^2, \;
(u\leftrightarrow v). \quad
\ea
These equations form the basis of the following analysis, and
hence a few remarks are in order.

First suppose that $\Lambda=0$. Then, given (\ref{WKK1}), $\alpha$
is harmonic and, as before, we can set $\alpha=\r$ without loss of
generality.  Note then that equations (\ref{WKK2})-(\ref{WKK3}) for
($\Omega,A)$ and (\ref{WKK4}) for the potentials $\Psi_\mu$
completely decouple. The former pair are analogous to the
Weyl-Papapetrou equations of (\ref{ZZ}) with, however, $\gamma$ and
$\delta$ arbitrary, whereas (\ref{WKK4}) are just Weyl potential
equations (\ref{weylrod}). The equations (\ref{WKK5}), which we
shall call integrability conditions, relate all potentials together
giving the function $\nu$. Thus, we have shown that, when
$\Lambda=0$, the $d$-dimensional system decouples to a
``Lewis-Papapetrou pair" on the one hand and $d-2$ Weyl potentials
on the other hand. Given the analysis of section \ref{sec:over}, for
$\Lambda=0$ and $d$ arbitrary, the  system involving a single
$A$-component  is therefore (essentially) integrable. Note that this
decoupling is a consequence of three facts: {\it i)} we have set
$\Lambda=0$; {\it ii)} there is only one non-zero angular momentum
and {\it iii)} the choice of our metric components (\ref{trans}).
Indeed, with the choice of (\ref{trans}) we can conveniently rewrite
the matrix of potentials \cite{Harmark} so that they are all
diagonal modulo the 2 by 2 matrix involving $(\Omega,A)$.

When $\Lambda\neq 0$, $\alpha$ is no longer harmonic and, hence, an
adapted coordinate system for $\alpha$ can no longer be chosen: this
is the major difficulty with the addition of the cosmological
constant. Now (\ref{WKK1}) gives $\nu$ in terms of $\alpha$ which
can then be substituted in (\ref{WKK5}) at the expense of raising
the order of the equation. Furthermore, the different potentials
(\ref{WKK2}-\ref{WKK4}) are coupled through $\alpha$. Despite this,
a number of symmetries can presently be identified and extended from
$\Lambda=0$ to $\Lambda\neq 0$, as we will now see.

Consider in particular the possible generalisation of EM duality
(\ref{EM}). Given the form of Maxwell's equation (\ref{WKK2}),
define a dual potential $\omega$ through \be \label{coffee}
(-\partial_z\omega, \partial_r\omega)=e^{\Omega}
\alpha^{\frac{d-4}{d-2}+\gamma\delta} (\partial_r A, \partial_z A).
\ee This is simply the analogous of the second equation of
(\ref{EM}), with $\omega$ the vector potential of the Hodge dual of
$F_{\mu \nu}$. In terms of $\omega$, the equations
(\ref{WKK2}-\ref{WKK3}) and (\ref{WKK5}) take the rather similar
form \ba \label{WKKD2} 0 &=& \overrightarrow{\nabla} \cdot \left (
e^{-\Omega}
  \alpha^{-\frac{d-4}{d-2}-\gamma \delta} \overrightarrow{\nabla} {\omega} \right )  \\
\label{WKKD3} \frac{1}{\alpha} \overrightarrow{\nabla} \cdot \left
( \alpha \overrightarrow{\nabla} {\Omega} \right ) &=&
\frac{\epsilon s}{2} e^{-\Omega} \alpha^{-\gamma\delta-\frac{2(d-3)}{d-2}} \left
(\overrightarrow{\nabla} \omega\right
  )^2 \\
\label{WKKD5} 2 \nu_{,u} \frac{\alpha_u}{\alpha}-
\frac{\alpha_{,uu}}{\alpha} &=& \frac{1}{s} \left(\Psi_{*,u}^2
+\frac{1}{2}  \Omega_{,u}^2 \right) -\frac{\epsilon}{2}
e^{-\Omega} \alpha^{-\gamma\delta -\frac{2(d-3)}{d-2}}  \left (
\omega_{\, ,u} \right )^2+\nonumber\\
&+& \sum_{{\hat\mu}\neq \mu_*}\Psi_{{\hat\mu},u}^2 , \qquad
(u\leftrightarrow v).
\ea
Comparing (\ref{WKK2}-\ref{WKK5}) with (\ref{WKKD2}-\ref{WKKD5}),
it is clear that, for $\Lambda=0$~{\footnote{Note that the parameter
    $\delta$ is redundant in the absence of a cosmological constant and can be
  set to any value. Here we take $\delta=0$.}} and $d=4$, one can associate to
every given solution of (\ref{WKK2}-\ref{WKK5})
a dual solution of these same equations through the map
\be
\label{EM1}
{\E\M} = \protect\left\{ \begin{array}{lll}
\Omega  \rightarrow  - \Omega \, ,
\\
A  \rightarrow  \omega \, ,
\\
\epsilon \rightarrow -\epsilon
\end{array} \protect\right. .
\ee
This is just the EM duality. For $\Lambda \neq 0$ this duality no
longer holds. Consider instead the map
\be
\label{duality}
{\A} = \protect\left\{ \begin{array}{lll}
\Omega  \rightarrow  \bar\Omega=- \Omega \, ,
\\
\omega  \rightarrow  \bar{A} = \omega \, ,
\\
\gamma \rightarrow  \bar{\gamma}=\pm \gamma
\\
\delta \rightarrow  \bar{\delta} = \mp \left[\delta
+\frac{2(d-4)}{(d-2)\gamma}\right]\, \\
\epsilon \rightarrow -\epsilon
\end{array} \protect\right. .
\ee It is straightforward to observe that any solution
$(\Omega,A,\Psi_\mu)$ of (\ref{WKK2}-\ref{WKK5}) with given values
of $\gamma\delta$ and $\epsilon$ gives rise, through (\ref{coffee}),
to a solution $(\Omega,\omega , \Psi_\mu)$ of the dual system
(\ref{WKKD2}-\ref{WKKD5}) with the same values of $\gamma\delta$ and
$\epsilon$  and that the latter can be mapped to a new solution of
(\ref{WKK2}-\ref{WKK5}) through $\A$. Unfortunately, the map $\A$
also alters (\ref{WKK1}) and therefore the symmetry is lost. There
is one exception and it occurs if and only if $d=4$: then $\A$
simply changes the sign of $\gamma$ or of $\delta$ leaving
(\ref{WKK1}) unaffected. As we will see in section \ref{sec:5}, a
consequence of this is that given a dilatonic electric solution with
$d=4$ and $\gamma \delta = -1$, the map $\A$ can be used to generate
a $D=5$ dimensional magnetic solution (and conversely). Finally,
note that whilst \be \label{embis} (\Omega, A, \Psi_\mu)
\stackrel{\mbox{(\ref{coffee})}}{\longrightarrow} (\Omega,\omega ,
\Psi_\mu) \stackrel{\A}{\longrightarrow} (\bar{\Omega}, \bar{A},
\Psi_\mu) \ee is built as an extension of the EM duality to $\Lambda
\neq 0$ for $d=4$, it is {\it not} the EM duality (\ref{EM}).
Indeed, the EM duality leaves unchanged the action parameter
$\gamma$ and exchanges solutions within the {\it same} theory,
i.e.~with the same dimension (here $d=4$) and the same parameter
$\gamma$. In contrast, the transformation (\ref{embis}) with $d=4$
and $\Lambda \neq 0$ exchanges solutions corresponding to {\it
different} theories, i.e.~with different parameters $\gamma$ and
$\delta$. In other words an uplifted $D=5$ rotating solution will
always be mapped to a $d=4$ EMD solution and not to a new $D=5$
solution.

\subsection{Ernst potentials with a cosmological
constant}\label{subsec:ernst}

We now proceed to generalise the method of Ernst to $\Lambda \neq
0$ and $d>3$. In analogy with (\ref{TPs}), let us define a complex
potential
\be \E_-=e^{\frac{\Omega}{2}}
\alpha^{\frac{\gamma\delta}{2}+\frac{d-3}{d-2}}+i\frac{\sqrt{s}}{2}
 \omega .
 \label{Em}
\ee Then, {\it in the electric field case only}, $\epsilon=-1$, the
Maxwell and scalar field equations (\ref{WKKD2}) and (\ref{WKKD3})
reduce to the single complex equation \be \label{zernst}
\frac{1}{\alpha} \del \cdot \left (\alpha \del \E_- \right ) =
\frac{(\del
  \E_-)^2}{\mbox{Re}(\E_-)} + \left(\frac{\gamma\delta}{2} +\frac{d-3}{d-2}\right)
  \mbox{Re}(\E_-) \frac{\Delta \alpha}{\alpha}.
\ee
For $\Lambda\neq 0$, $\alpha$ is not a harmonic function and
therefore there is an extra term in the Ernst equation relative to
its original form (\ref{E4}).

In the magnetic field case, $\epsilon=+1$, equations (\ref{WKKD2})
and (\ref{WKKD3}) can no longer be written in such an Ernst form.
However, one can return to the system (\ref{WKK2}) and (\ref{WKK3}):
{\it in the magnetic field case $\epsilon=+1$ only}, these may be
derived from the potential \be \label{Eplus}
\E_+=e^{\frac{-\Omega}{2}}
\alpha^{-\frac{\gamma\delta}{2}+\frac{1}{d-2}}+i \frac{\sqrt{s}}{2}
A \ee with corresponding equation \be \label{zernstM}
\frac{1}{\alpha} \del \cdot \left (\alpha \del \E_+ \right ) =
\frac{(\del
  \E_+)^2}{\mbox{Re}(\E_+)} + \left(-\frac{\gamma\delta}{2} +\frac{1}{d-2}\right)
  \mbox{Re}(\E_+) \frac{\Delta \alpha}{\alpha}.
\ee

Let us now consider the cases where the last term on the
RHS of (\ref{zernst}) or(\ref{zernstM}) vanishes. The electric or
magnetic Ernst equation then reduces
to the standard one (\ref{E4}), however with the difference that
$\alpha$ is not harmonic. Furthermore, in these cases,  we note that
(\ref{WKK4}) or (\ref{zernstM}) can be derived from the
two-dimensional action \be \label{sigmamodelaction} S_2 = \int dr \,
dz \, \alpha(r,z) \left [ \frac{\del \E \cdot \del \E^*}{(\E +
\E^*)^2} +
 \sum_{\mu=0}^{d-3} \left ( \del \Psi_\mu \right )^2
\right ] \, . \ee where $\E$ stands for either $\E_-$ or $\E_+$.
Thus we have ended up with a non-linear $\sigma$-model, whose target
space is spanned by the coordinates $(\E, \E^*, \Psi_\mu)$ and is
endowed with the $d$-dimensional metric \be \label{metricsigma}
G_{d} = \frac{d\E \, d\E^*}{(\E + \E^*)^2} + \sum_{\mu=0}^{d-3}
\left (d\Psi_\mu \right )^2 = \frac{d\xi \,
d\xi^*}{(1-\left|\xi\right|)^2}+ \sum_{\mu=0}^{d-3} \left (d\Psi_\mu
\right )^2 \, , \ee where, as in (\ref{xidef}), we have set \be
\label{ernstxi} \E \equiv \frac{\xi - 1}{\xi +1} \, . \ee The target
space is thus a $d$-dimensional Riemannian manifold which is locally
isometric to ${\mathbb H}_2 \times {\mathbb R}^{d-2}$, where
${\mathbb H}_2$ is the hyperbolic plane.

This symmetry can be exploited only if it is respected by the integrability condition,
(\ref{WKK5}) or (\ref{WKKD5}). For both eletric and magnetic cases, this is possible only
if $s=4$. Indeed, in this case, the integrability condition can be conveniently rewritten in terms of  $(\E,
\E^*, \Psi_\mu)$, as
\be
\label{intcondernst}
2 \nu_{,u} \frac{\alpha_{,u}}{\alpha}- \frac{\alpha_{,uu}}{\alpha} =
 2 \, \frac{\E_{,u} \,  \E_{,u}^*}{\left (\E + \E^* \right )^2} +
\frac{1}{4}\Psi_{*,u}^2 +  \sum_{i=1}^{d-3}\Psi_{i,u}^2 \ee where,
up to a renormalisation of the $\Psi_\mu$ fields, we recognise on
the RHS the target space metric $G_d$. Since the fields $(\E, \E^*,
\Psi_\mu)$ only enter the field equations through $G_d$, each
transformation of the target space isometry group leaves the field
equations invariant. For example, the transformation \be \label{apc}
\forall \vartheta \in {\mathbb R}, \qquad \xi \rightarrow
e^{i\vartheta} \xi \ee is clearly such an isometry and for each
constant phase $\vartheta$ will yield a different solution. Thus we
can generate different solutions of the field equations through the
action of the universal cover $SU(1,1) \times E_{d-2}$ of the
isometry group $SO(2,1) \times E_{d-2}$ of the target space.

It is interesting to reflect on a geometric interpretation of the
field equations (\ref{WKK1}-\ref{WKK5}), or (\ref{WKKD2}-\ref{WKKD5}). Note for a start the volume
element $d\r\,d\z\,\alpha$ appearing in (\ref{sigmamodelaction}).
For $\Lambda=0$, in (\ref{sigmamodelaction}) the manifold over which
integration takes place is the 3-dimensional flat cylindrical
metric. When $\Lambda\neq 0$, on the other hand, the metric is still
axially symmetric but is no longer flat \be \label{hood}
d\r^2+d\z^2+\alpha(\r,\z)^2 d\varphi^2. \ee It is intriguing to note
that the scalar curvature of (\ref{hood}) is given by the component
$e^{2\nu}$ via equation (\ref{WKK1}) and this, in turn, says that in
the presence of the cosmological constant, (\ref{hood}) is a curved
metric whose curvature depends on $\E$ and $\Psi_\mu$. Actually,
the LHS differential operators acting on $\E$ and $\Psi_\mu$ in
(\ref{WKK2}-\ref{WKK4}), or (\ref{WKKD2}-\ref{WKKD5}), are the Laplace operators associated to the
metric $(\ref{hood})$. In some sense, the integrability condition
(\ref{WKK5}), or (\ref{WKKD5}), can be seen to relate the 'geometry', on the LHS, to
'matter', on the RHS of the field equations. This geometric
interpretation is another way to approach the field equations that
deserves future study.

For the magnetic case, the two conditions on $\gamma$ and $\delta$ discussed above are equivalent  to
(\ref{gd}), which corresponds to the case where the $d$-dimensional system can be uplifted to
a $D$-dimensional solution.
For the electric case, the two conditions on the couplings
are
\be
\label{gdelecernst}
\gamma=\pm \sqrt{\frac{2(d-1)}{d-2}}, \qquad
\gamma \delta = -2 \frac{d-3}{d-2} \, ,
\ee
where the first condition ensures that the integrability equation
(\ref{WKKD5}) can be written in the form (\ref{intcondernst}) with,
whereas the second condition ensures
that the last term on the RHS of (\ref{zernst}) vanishes. In
 the particular case $d=4$, it is
possible, using the map $\A$,  to relate an electric EMD
$d=4$-dimensional solution to a magnetic $5$-dimensional solution.
This is an interesting way to lift dilatonic electric solutions to
$5$ dimensions.
 To summarize, we have the following diagram
$$\xymatrix{\gamma\delta =1 \qquad & (g_5)_{\mbox{\scriptsize Magn.}}
  \ar[rr]^-{SU(1,1)} \ar[d]^-{{\A}}
&& (g_5')_{\mbox{\scriptsize Magn.}}  \ar[d]^-{{\A}} \\
\gamma\delta = -1 \qquad & (g_4 \oplus A \oplus \Phi)_{\mbox{\scriptsize
  Elec.}} \ar[rr]^-{SU(1,1)}
&& (g_4' \oplus A' \oplus \Phi')_{\mbox{\scriptsize Elec.}}  }$$

\subsection{Extending the Papapetrou method}\label{subsec:papa}

We now consider the generalisation of a construction technique
of Papapetrou \cite{Papa1} which was originally carried out in 4
dimensions with $\Lambda=0$ (see section \ref{sec:over}). Here
we consider the general case of a $d$-dimensional EMD system with
$\Lambda \neq 0$. We will show that when the real potentials
$\Omega$ and $\Psi_\mu$ are functionals of
the EM potential $A$ or $\omega$, the
$d$-dimensional EMD system
reduces to a Weyl system with $\Lambda\neq 0$ {\it provided}
certain constraints on the coupling constants $\gamma$ and
$\delta$ are satisfied.

We will consider simultaneously the two cases $\Omega=\Omega(A)$
and $\Omega=\Omega(\omega)$ and write generically
$\Omega=\Omega(X)$ with $X=A,\omega$. In both cases, the equations
(\ref{WKK2}-\ref{WKK3}) and (\ref{WKKD2}-\ref{WKKD3}) reduce to
\ba
\left[ \Delta X + \frac{\del \alpha}{\alpha} \cdot \del X  \right] + \q \Omega'(\del
X)^2 &=& 0 \label{Pa1}
\\
\Omega' \left[ \Delta X +  \frac{\del \alpha}{\alpha} \cdot \del
X \right] + \left\{ \Omega'' - \frac{\epsilon s}{2}     e^{\q \Omega} \right\} (\del X)^2 &=& 0 \label{Pa2}.
\ea
where a prime denotes an ordinary derivative with respect to $X$, provided

\be \label{bucher1}  \protect\left\{\begin{array}{ll}  \q=1, \quad
\gamma \delta=
    \frac{2}{d-2},\quad \mbox{for} \quad X=A
\\
\q=-1, \quad \gamma \delta= -2\frac{d-3}{d-2}, \quad \mbox{for}
\quad X=\omega.
\\
\end{array}\protect\right.
\ee
The conditions on the couplings, which are the same as those
encountered in the previous subsection, are necessary to get the
same expression in the brackets on the left hand side of
(\ref{Pa1}) and (\ref{Pa2}).  Taking the difference we get the
ordinary differential equation
\be
\Omega'' -  \frac{s \epsilon}{2}  e^{\q \Omega} - \q(\Omega')^2 = 0
\label{AOmega}
\ee
with solution
\be
e^{-\q\Omega} =\left (-\frac{\epsilon \q s}{4}  X^2 + k_1 X + k_0
\right ) \, , \label{om}
\ee
where $k_1$ and $k_0$ are some integration constants.
The same trick can be used in (\ref{WKK4}) for each $\Psi_\mu$, once we let
$\Psi_\mu=\Psi_\mu(X)$. The  solution reads
\be
\Psi'_\mu = l_\mu \, e^{\q\Omega} \, ,
\ee
where the $l_\mu$'s are again constants of integration.

It is now convenient to introduce the function
\be
\label{varum} \varphi(X)=\sqrt{2\lambda} \int^X
\frac{dx}{-(\epsilon \q s/4) x^2 + k_1 x + k_0} \, ,
\ee
so that, using (\ref{om}),  $\varphi_u^2 = 2 \lambda e^{2\q\Omega}
(X_u)^2$. $\lambda$ is a free constant which we now fix by taking
into account the last equations --- the integrability conditions
(\ref{WKK5}) and (\ref{WKKD5}) --- which become
\ba
2 \nu_{,u} \frac{\alpha_u}{\alpha}- \frac{\alpha_{,uu}}{\alpha}
&=& \left( \frac{\varphi_u^2}{2 s\lambda} \right) \left\{ l_0^2 +
\left[
\frac{X^2}{8}(s^2-16) - \frac{2X\epsilon \q k_1}{s}(s-4) \right] \right. \nonumber \\
&& \qquad \qquad \qquad \; \; \; + \left. \left( \frac{k_1^2}{2} +
\frac{8 \epsilon \q k_0}{s}\right) + \sum_{i=1}^{D-4} l_i^2
 \right\}\ .
\ea Requiring that the RHS of the above equation be independent of
$X$ yields $s=4$ or, according to (\ref{s}), \be \gamma = \pm
\sqrt{\frac{2(d-1)}{(d-2)} }. \label{gammaf} \ee  In this case \be
\label{dav} \varphi(X) =-\q\epsilon \sqrt{2\lambda} \protect\left\{
\begin{array}{lll} \frac{1}{\sqrt{k_1^2+4 \q\epsilon k_0}}\ln \left(
\frac{X-(\q\epsilon k_1/2) - \sqrt{(k_1^2/4)+ \q\epsilon k_0}}
{X-(\q\epsilon k_1/2) + \sqrt{(k_1^2/4)+ \q\epsilon k_0}}
\right)+c_0, \qquad \qquad \; \; k_1^2>-4 \q\epsilon k_0 \nonumber
\\
\\
-\frac{1}{X-\q\epsilon k_1/2}+c_1, \qquad \qquad \qquad \qquad
\qquad \qquad \qquad \qquad \; \;
k_1^2=-4\q\epsilon  k_0 \nonumber \\
\\
\frac{1}{\sqrt{-\q\epsilon k_0-k_1^2/4}} \arctan \left(
\frac{X-\q\epsilon k_1/2}{\sqrt{-\q\epsilon k_0-k_1^2/4}}
\right)+c_2,
\qquad \qquad \qquad k_1^2<-4 \q\epsilon k_0 \\
\end{array} \protect\right.
\ee where $c_0$, $c_1$ and $c_2$ are integration constants. Then on
choosing \be \lambda = \frac{1}{4} \left\{l_0^2 + \left(
\frac{k_1^2}{2} + 2 \epsilon \q k_0 \right) + \sum_{i=1}^{D-4} l_i^2
\right\}, \ee the integrability equations (\ref{WKK5}) or
(\ref{WKKD5}) reduce to \beq 2 \nu_{,u} \frac{\alpha_{,u}}{\alpha}-
\frac{\alpha_{,uu}}{\alpha} = \frac{1}{2} \varphi_{,u}^2, \qquad
(u\leftrightarrow v). \eeq

When $\Omega=\Omega(A)$, the conditions on $\gamma$ and $\delta$,
(\ref{bucher1}) and (\ref{gammaf}), are equivalent to (\ref{gd}),
{\it i.e.} to an uplifted $D$ dimensional rotating spacetime.
Thus, as stated earlier, under the hypothesis (\ref{bucher1}) and
(\ref{gammaf}), each $D=d+1$-dimensional Weyl solution of
Einstein's equations yields a family of $D$-dimensional stationary
and axisymmetric solutions. Indeed, the field equations
(\ref{WKK2})-(\ref{WKK5}) reduce to
\ba
\label{W1}
\Delta \alpha &=& -2 \Lambda \alpha^{\frac{1}{D-2}} e^{2\nu}, \\
\label{W2} {\del} \cdot \left ( \alpha
 \del {\varphi} \right ) &=&  0,\\
\label{W3} 2 \nu_{,u} \frac{\alpha_{,u}}{\alpha}-
\frac{\alpha_{,uu}}{\alpha} &=& \frac{1}{2} \varphi_{,u}^2, \qquad
(u\leftrightarrow v). \ea The Weyl metric element is (here we take
$\epsilon=-1$) \be ds^2 = e^{2\nu}
\alpha^{-\frac{D-3}{D-2}}(d\r^2+d\z^2)+\alpha^{2\over {D-2}}
\left[-e^{\sqrt{\frac{2(D-3)}{(D-2)}}\varphi} dt^2 +
e^{-\sqrt{\frac{2}{(D-2)(D-3)}}\varphi}
  \sum_{i=1}^{D-3} (dx^i)^2\right] \, ,
  \label{tofv}
\ee
where $\varphi$ is given in (\ref{varum}) with $\q=1$, $X=A$ and
$s=4$. Note that even if $D>4$ we have only a single Weyl field
$\varphi$ in (\ref{W1}-\ref{W3}) since we have assumed a single
angular momentum component $A$ in (\ref{dav}). The metric solutions
obtained this way have a very particular form. Indeed, using
(\ref{om}) and (\ref{varum}) (see also (\ref{melina})), we find that
a rotating spacetime metric reduces to \ba \label{melina1} ds^2 &=&
e^{2\nu} \alpha^{-\frac{D-3}{D-2}}(d\r^2+d\z^2)+\alpha^{2\over
{D-2}}
\left[\frac{e^{-\sqrt{\frac{D-4}{2(D-2)}}\Psi_*}}{\sqrt{A^2+k_1
A+k_0}}\left( -dt^2-2 A d\varphi dt \right. \right. \nonumber
\\
&   & \qquad +  \left. \left. (k_1 A + k_0)d\varphi^2 \right) +
e^{\sqrt{\frac{2}{(D-2)(D-4)}}\Psi_*}
 \sum_{i=1}^{D-4}e^{2\Psi_i}(dx^i)^2\right] \, ,
\ea
with
\be
\label{pap1} \Psi_\mu=\frac{l_\mu}{\sqrt{2\lambda}}\varphi.
\ee

When, in turn, $\Omega=\Omega(\omega)$, then (\ref{bucher1}) and
(\ref{gammaf}) give
\be
\gamma = \pm \sqrt{ \frac{2(d-1)}{(d-2)} } \qquad
\delta = \mp(d-3) \sqrt{ \frac{2}{(d-2)(d-1)} } \, ,
\ee
which is {\it not} equivalent to a $D$-dimensional system but is a
particular EMD $d$ dimensional system. The duality of the previous
section, however, tells us that when $d=4$ in particular we will
be able to map any Papapetrou solution  to a $D=5$ rotating
spacetime solution. For the dual system, the field equations
reduce to
\ba
\label{WD1}
\Delta \alpha &=& -2 \Lambda \alpha^{-\frac{(d-5)}{(d-1)}} e^{2\nu}, \\
\label{WD2} \overrightarrow{\nabla} \cdot \left ( \alpha
  \overrightarrow{\nabla} {\varphi} \right ) &=&  0,\\
\label{WD3} 2 \nu_{,u} \frac{\alpha_{,u}}{\alpha}-
\frac{\alpha_{,uu}}{\alpha} &=& \frac{1}{2} \varphi_{,u}^2, \qquad
(u\leftrightarrow v) \ea where $\varphi$ is given in (\ref{varum})
with $\q=-1$, $X=\omega$, $s=4$ (see in particular \cite{christos}).
Note that for $d=4$ equations (\ref{WD1}-\ref{WD3}) are identical to
(\ref{W1}-\ref{W3}) for $D=5$ in agreement with the duality map $\A$.

Indeed, as
discussed in section \ref{sec:over}, Papapetrou's construction was
originally carried out for the dual system and then mapped in
$D=4$ dimensions via EM duality. In other words, one supposes
rather that $\Omega=\Omega(\omega)$ and evaluates $A$
independently from (\ref{coffee}). In crude terms, this means that
the rotation field $A$ will generically depend on a different
coordinate from $\Omega$ and the metric will not be of the
specific form (\ref{melina1}). In the absence of a cosmological
constant we can apply the same method in arbitrary dimensions:
when $\Lambda\neq 0$, however, we can only do so for $D=5$.

In \cite{Charmousis:2003wm} it was shown that the system
(\ref{W1})-(\ref{W3}) is completely integrable if one makes the
hypothesis that $\varphi$ depends only on one of the two
coordinates, say $\z$.    It then follows that the canonical
components $A,\Omega,\Psi_\mu$ must also depend the same variable
$\z$.  Furthermore, from (\ref{W2}), $\alpha$ is separable: \be
\alpha=f(\r)g(\z), \qquad g(\z)=\frac{c}{\varphi_{,\z}} \ee where
$c$ is a nonvanishing constant if $\varphi_{,\z}\neq 0$. The
remaining two equations (\ref{W1}) and (\ref{W3}) then give $f(\r)$
and $g(\z)$. As was discussed in \cite{Charmousis:2003wm}, there are
three classes of possible solutions: class I with $f_{,\r} =0$,
class II with $g_{,\z} =0$, and class III with both $f_{,\r},
g_{,\z} \neq 0$.  We will return to these in sections \ref{sec:4}
and \ref{sec:5} where we discuss solutions in $D=4,5$ dimensions.
Finally, note that the same method also gives a large class of
solutions to the dual system given in (\ref{WD1}-\ref{WD3}).

\subsection{Set-up for uplifted spacetimes in $D$
dimensions}\label{subsec:Dcon}
In this section, we focus on $D$-dimensional solutions which can be
obtained from uplifting $d=D-1$ dimensional EMD solutions. We start
by summarizing our results in this specific case, corresponding to
values of $\gamma$ and $\delta$ given in (\ref{gd}). From
(\ref{KKmetric}), (\ref{Weyll1}) and (\ref{trans}), the metric in
the electric case (following a Wick rotation $x_0 \rightarrow i
\varphi$ and $w \rightarrow i t$) corresponds to a rotating metric,
\ba \label{melina} ds^2 &=& e^{2\nu}
\alpha^{-\frac{D-3}{D-2}}(d\r^2+d\z^2)+\alpha^{2\over
  {D-2}}
\left[e^{-\sqrt{\frac{D-4}{2(D-2)}}\Psi_*}\left[ -e^{\Omega\over
2}(dt+A
  d\varphi)^2+e^{-\Omega\over 2}
  d\varphi^2\right]\right.+ \nonumber \\
 &&+ \left.e^{\sqrt{\frac{2}{(D-2)(D-4)}}\Psi_*}
  \sum_{i=1}^{D-4}e^{2\Psi_i}(dx^i)^2\right] \, .
\ea The pole at $D=4$ is artificial since then the $\Psi_\mu=0$.
After an analytic continuation of the time coordinate $x_0
\rightarrow i x_{D-4}$, the magnetic spacetime is given by \ba
\label{luna} ds^2 &=& e^{2\nu}
\alpha^{-\frac{D-3}{D-2}}(d\r^2+d\z^2)+\alpha^{2\over
  {D-2}}
\left[e^{-\sqrt{\frac{D-4}{2(D-2)}}\Psi_*}\left[ e^{\Omega\over
2}(dw+A
  d\varphi)^2+e^{-\Omega\over 2}
  d\varphi^2\right]\right.+ \nonumber \\
 &&+ \left.e^{\sqrt{\frac{2}{(D-2)(D-4)}}\Psi_*}
  \sum_{i=1}^{D-4} e^{2\Psi_i}(dx^i)^2\right],
\ea which is a purely Riemannian.

From (\ref{WKK1})-(\ref{WKK5}), the field equations take the
rather simplified form
\ba
\label{PD1}
\Delta \alpha &=& -2 \Lambda \alpha^{\frac{1}{D-2}} e^{2\nu}, \\
\label{PD2}
0 &=& \overrightarrow{\nabla} \cdot \left ( e^{\Omega}
  \alpha \overrightarrow{\nabla} {A} \right )  ,\\
\label{PD3} \frac{1}{\alpha} \overrightarrow{\nabla} \cdot \left (
\alpha \overrightarrow{\nabla} {\Omega} \right ) &=& 2 \,
\epsilon \, e^{\Omega} \left (\overrightarrow{\nabla} A \right )^2 ,\\
\label{PD4}
\overrightarrow{\nabla} \cdot \left ( \alpha
  \overrightarrow{\nabla} {\Psi_\mu} \right ) &=&  0
,\qquad \mu=0...d-3
\\
\label{PD5} 2 \nu_{,u} \frac{\alpha_u}{\alpha}-
\frac{\alpha_{,uu}}{\alpha} &=& \frac{1}{4} \left(\Psi_{*,u}^2
+\frac{1}{2}  \Omega_{,u}^2 \right) +\frac{\epsilon}{2} e^{\Omega}
\left ( A_{\, ,u} \right )^2 + \sum_{i=1}^{D-4}\Psi_{i,u}^2, \qquad
(u\leftrightarrow v). \ea The electric Ernst potential (\ref{Em})
now reads \be \E_-=e^{\frac{\Omega}{2}}\alpha+i\omega,
 \label{Em1}
\ee replacing (\ref{PD2}-\ref{PD3}) by \be \label{zernstD}
\frac{1}{\alpha} \del \cdot \left (\alpha \del \E_- \right ) =
\frac{(\del \E_-)^2}{\mbox{Re}(\E_-)} + \mbox{Re}(\E_-) \frac{\Delta
\alpha}{\alpha}. \ee In the magnetic case, $\epsilon=+1$,
(\ref{Eplus}) becomes \be \E_+=e^{-\frac{\Omega}{2}}+i A
\label{zzzzM} \ee with corresponding equation \be \label{zernstMD}
\frac{1}{\alpha} \del \cdot \left (\alpha \del \E_+ \right ) =
\frac{(\del
  \E_+)^2}{\mbox{Re}(\E_+)}
\ee
where, as we have already noted, the extra term of (\ref{zernstM})
drops out.

A particular class of solutions can be found taking advantage of
the `decoupled' form of the field equations (\ref{PD1}-\ref{PD5}).
Indeed, suppose that $\alpha$ and $\nu$ only depend on $\r$
whereas $\Omega$, $\Psi_\mu$ and $A$ only depend on $\z$. In that
case the equations decouple into two separate systems of ODEs; one
$\r$-dependent for $\alpha$ and $\nu$; and one $\z$-dependent for
the remaining fields. Following the geometric interpretation of
section 3 this amounts to splitting contributions from geometry
and matter and treating them separately. The $\r$-dependent system
reads
\ba
\alpha'' &=& -2\Lambda \alpha^{\frac{1}{D-2}} e^{2\nu} \\
2\nu' \frac{\alpha'}{\alpha} &=& \frac{\alpha''}{\alpha} \, ,
\ea
where a prime stands for a derivative
with respect to the unique variable $\r$. The system here is
identical to the one appearing in \cite{bowcock,tony} and the
solution reads
\ba
e^{2\nu} &=& \alpha' \\
\label{coc} \alpha' &=& -\frac{\mu}{(D-2)^2} - \frac{2
(D-2)\Lambda}{(D-1)} \alpha^{\frac{D-1}{D-2}} , \ea
where $\mu$ is some real integration constant. In
\cite{tony} if $\z$ is a spacelike coordinate then the solutions
of (\ref{coc}) can be coordinate transformed to an adS soliton
\cite{mers}. On the other hand, if $\z$ is a timelike coordinate
\cite{bowcock} one gets an adS planar black hole (see section 5).
So we anticipate to recover these two solutions as a special case
and furthermore to obtain continuous deformations of these. We
will present these in detail in Section 5 for $D=5$ dimensions.

Observe that all equations are independent of $D$ except (\ref{PD1})
when $\Lambda\neq 0$. The metrics (\ref{melina}) or (\ref{luna}),
however, themselves depend on the dimension. Therefore, the form of
the $D$-dimensional field equations dictates an important result:
for $\Lambda =0$ and given a $D$-dimensional solution, we can always
construct a higher dimensional $D+n$ ($n$ positive integer)
dimensional solution. Indeed, recall first that when $\Lambda=0$,
$\alpha$ can be taken as the radial coordinate. Now, suppose one
takes a known $D$ dimensional solution $(\Omega,A,\Psi_\mu)$, where
$\mu=0, \ldots, D-4$.  Then, a new $D+n$ solution
$(\Omega,A,\Psi_\nu)$, for $\nu=0,\ldots,D+n-4$, can be obtained
from the $D$-dimensional solution simply by calculating the new Weyl
potentials from (\ref{PD4}), so that $\nu_{D+n}$ is given by direct
integration of $(\ref{PD5})$. That this is a new solution of the
Einstein equations is due to the fact that (\ref{PD5}) relates the
different potentials together independently of the spacetime
dimension.  To summarise, taking an arbitrary stationary and
axisymmetric solution in 4 dimensions, such as Kerr or TN say, we
can construct higher dimensional solutions by adding $n$ extra Weyl
potentials. Unfortunately this property is spoiled once we switch on
$\Lambda$, since from (\ref{PD1}), the component $\nu$ becomes a $D$
dependent quantity and $\alpha$ is no longer free.

Conversely, for $\Lambda=0$, a higher dimensional stationary
solution of axial symmetry with one angular momentum will always
originate from a unique 4 dimensional seed solution {\it with the
same Ernst potentials} $\E_{\pm}$. Consider, for example, a known
$D+1$ dimensional solution and let us look for the $D$-dimensional
seed solution. The only unknown metric component is $\nu_D$ which is
immediately given from direct integration by (\ref{PD5}), \be
\label{basic} 2 (\nu_{(D),u}-
\nu_{(D+1),u})\frac{\alpha_{,u}}{\alpha}= -\frac{1}{4} \Psi_{,u}^2,
\qquad (u\leftrightarrow v) \, . \ee It will be be useful for
applications to define $\sigma=\nu_{(D),u}- \nu_{(D+1),u}$ and to
rewrite the above equation in terms of $\r$ and $\z$: \bea
\sigma_{,\z}=\frac{\alpha}{8 (\alpha_{,\z}^2
+\alpha_{,\r}^2)}\left[\alpha_{,\z}
  ({\Psi}_{,\r}^2-{\Psi}_{,\z}^2)-2\alpha_{,\r}{\Psi}_{,\z}{\Psi}_{,\r} \right] \, ,\nonumber\\
\label{gif} \sigma_{,\r}=-\frac{\alpha}{8(\alpha_{,\z}^2
+\alpha_{,\r}^2)}\left[\alpha_{,\r}
  ({\Psi}_{,\r}^2-{\Psi}_{,\z}^2)+2\alpha_{,\z}{\Psi}_{,\z}{\Psi}_{,\r} \right] \, .
\eea Note that these equations are particularly simple in Weyl
coordinates. Simple examples of this and of previous methods will be
given in the following sections.

\section{Examples in $D=4$ dimensions}\label{sec:4}

The aim of this section is two fold. First, we make the connection
between our general analysis of section \ref{sect:Dim} and the well known results of
$D=4$ and $\Lambda=0$  general relativity (as summarised briefly in section \ref{sec:over}) .
Second, we give
examples of Ernst potentials for well-known GR solutions, though now
extended to the case of spacetimes with non-zero cosmological constant, $\Lambda \neq 0$.

Our general starting point is the electric EMD system
(\ref{WKK1}-\ref{WKK5}) which reads for $d=3$  \ba \label{4d1}
\Delta \alpha &=& -2\Lambda \alpha^{1-\frac{\delta^2}{2}} e^{2\nu}, \\
\label{4d2}
\del \cdot \left (\frac{e^{\Omega}}{\alpha^{1-\gamma\delta}} \del A \right ) &=&0, \\
\label{4d3}
\frac{1}{\alpha} \del \cdot \left (\alpha \del \Omega \right ) +
\frac{\gamma^2}{2\alpha^{2-\gamma\delta}} e^{\Omega} (\del A)^2 &=& 0 ,\\
\label{4d4} \Delta \nu  + \frac{1}{4\gamma^2} (\del \Omega)^2 +
\frac{1-\gamma\delta}{4\alpha^{2-\gamma\delta}}e^{\Omega} (\del
A)^2
&=& \frac{1}{2} \left (1- \frac{\delta^2}{2} \right )\frac{\Delta \alpha}{\alpha}, \\
\label{4d5} 2\nu_{,u} \frac{\alpha_{,u}}{\alpha} -
\frac{\alpha_{,uu}}{\alpha} &=& \frac{1}{2\gamma^2} (\Omega_{,u})^2
- \frac{1}{2\alpha^{2-\gamma\delta}} e^{\Omega} A_{,u}^2 \quad (u
\leftrightarrow v) \ea From (\ref{gd}), for the special values of
the coupling constants namely $\gamma=2$ and $\delta=1$, we can
uplift to a $D=4$ dimensional axisymmetric and stationary spacetime.
Using (\ref{melina}), the metric in the above components reads \be
\label{4d} ds^2=e^{2\nu}\alpha^{-1/2}(d\r^2+d\z^2)+\alpha
e^{-\frac{\Omega}{2}}d\varphi^2-\alpha e^{\frac{\Omega}{2}} (dt+A
d\varphi)^2. \ee Note that the metric components differ from the
original Weyl-Papapetrou ones (\ref{stat}). Indeed, the Weyl
potential $\lambda$ is now given by $e^{2\lambda}=\alpha
e^{\frac{\Omega}{2}}$, although $\lambda$ and $\Omega$ obey a
similar differential equation (compare (\ref{ZZ}) and (\ref{4d3})).
Furthermore, when $\Lambda=0$ the component $\alpha$ is harmonic and
is the radial coordinate $\r$ in (\ref{stat}). These slight
differences are important, and result from having chosen variables
which absorb the cosmological constant term in the field equations
(\ref{4d1}-\ref{4d5}).

In the
magnetic case, the 4 dimensional metric is of Euclidean signature
and corresponds generically to a Euclidean instanton solution
\be
\label{4dd}
ds^2=e^{2\nu}\alpha^{-1/2}(d\r^2+d\z^2)+\alpha
e^{-\frac{\Omega}{2}}d\varphi^2+\alpha e^{\frac{\Omega}{2}} (dw+A
d\varphi)^2.
\ee

As discussed in section \ref{sect:Dim}, given the absence of an EM duality
transformation (\ref{duality})  when $\Lambda \neq 0$ we can define two different Ernst
potentials $\E_\pm$; $\E_-(\omega)$ given in (\ref{Em1}) for the electric
spacetime (\ref{4d}), and $\E_+(A)$ given in (\ref{zzzzM}) for a  magnetic spacetime. The
electric potential $\E_-$ is identical to the original
Ernst potential (\ref{E4}) for the metric (\ref{4d}).
As was discussed in section \ref{sec:over}, the electric Ernst potential and corresponding Ernst equation
were used in \cite{Ernst,piotr} to generate new solutions for $\Lambda=0$ such as,
for example, Schwarzschild spacetime using  spheroidal coordinates (\ref{prolate}).
Here, lacking a relevant coordinate system for $\Lambda \neq 0$, we merely construct the relevant potentials
for some well known solutions.

Consider first Carter's metric \cite{carter} which describes a rotating Kerr
black hole in an asymptotically adS spacetime:
\bea
\label{carter}
ds_4^2&=&-\frac{\Delta}{\rho^2}\left(dt-\frac{a\sin^2\theta}{\Xi_a}d\varphi\right)^2
+\frac{\Delta_\theta\sin^2\theta}{\rho^2}\left(adt-\frac{r^2+a^2}{\Xi_a}d\varphi\right)^2\nonumber\\
&+& \rho^2 \left({dr^2\over \Delta
  }+\frac{d\theta^2}{\Delta_\theta}\right),
\eea
where $k$ is the curvature scale of adS, $M$ is the black hole mass, $a$ the
angular momentum parameter and
\bea
\Delta&=&(r^2+a^2)(1+k^2 r^2)-2M r,\\
\Delta_\theta&=&1-a^2k^2\cos^2\theta,\quad\Xi_a=1-a^2k^2,\\
\rho^2&=&r^2+a^2\cos^2\theta,\quad\Lambda=-3k^2.
\eea
As a general rule, metrics with a cosmological constant cannot be written explicitly
in the coordinate system chosen
in (\ref{4d}).  However, this is not a problem since we can transit to the coordinate system
of (\ref{carter}) by setting
\be
{dr^2\over \Delta }=d\r^2, \qquad \frac{d\theta^2}{\Delta_\theta}=d\z^2,
\ee
meaning that $\z$ and $\r$ are implicitly given as functions or $\theta$ and
$r$, respectively.
Using (\ref{4d}), this is all we need to know in order to identify the different components:
\bea
\alpha &=& \frac{\sin\theta}{\Xi_a  }
\sqrt{\Delta\Delta_\theta},\\
A&=& {a\sin^2 \theta (\Delta-\Delta_\theta(r^2+a^2))\over {\Xi_a
    (a^2\Delta_\theta \sin^2 \theta-\Delta)}},\\
e^{\Omega}&=&\frac{\Xi_a^2 (\Delta - a^2\Delta_\theta \sin^2
  \theta)^2}{\Delta\Delta_\theta \rho^4 \sin^2 \theta }, \\
\e^{2\nu}&=& \rho^2 \alpha^{1/2}. \eea Using (\ref{Em1}) and
(\ref{coffee}) one finds that the electric Ernst potential for
Carter's solution is given by \be \label{brandon}
\E_-=\frac{1}{\rho^2}\left(\Delta-a^2\sin^2\theta \Delta_\theta - 2
i a \cos \theta ( k^2\rho^2 r +M)\right). \ee If there is no
rotation, $a=0$, the Ernst potential is real and corresponds to
Kottler's black hole \cite{kottler}. For $M=0$ we have pure adS but
the potential is still complex since the metric has non-zero angular
momentum{\footnote{This is quite unlike the situation for Kerr's
solution at asymptotic infinity.}}. If $\Lambda=0$, $\E_{-}$ is the
usual Ernst potential in the coordinates of (\ref{carter}).
Considering $\alpha=i a$ and $t=-iw$ we obtain the magnetic Carter
instanton (see \cite{D93}). The corresponding magnetic Ernst
potential is, according to (\ref{zzzzM}), \be
\E_+=\frac{\sin\theta}{\Xi_\alpha (\Delta + \alpha^2\Delta_\theta
\sin^2
  \theta)}\left(\sqrt{\Delta\Delta_\theta} \rho^2 - i \alpha\sin\theta (\Delta-\Delta_\theta(r^2-\alpha^2))\right).
\ee

Another interesting example is Taub-NUT spacetime with a $\Lambda$
term \cite{TN,carter} \be \label{TN4} ds^2= -F(r)(dt+A
d\varphi)^2+\frac{dr^2}{F(r)}+(r^2+n^2) d\Omega_{II}^2 \ee with \bea
A&=&2n\cos \theta, \nonumber \\
\label{bob}
F(r)&=&\frac{1}{l^2(n^2+r^2)}\left[r^4+(l^2+6n^2)r^2-2Mrl^2-n^2(l^2-3n^2)
\right]. \eea The constants $M$, $l$ and $n$ are the mass, the
length scale $l=1/k$ and the Taub-NUT parameter, respectively. The
electric Ernst potential is simply \be \E_-=F(r)+i\omega(r) \ee with
\be \label{GHoff}
\omega=-\frac{2n}{l^2(n^2+r^2)}(r^3-rl^2+Ml^2)-\frac{6n^2}{l^2}
\arctan(r/n), \ee whereas the magnetic potential is given by \be
\E_+=\frac{\sin\theta\sqrt{r^2-\n^2}}{\sqrt{F(r)}}+2i\n \cos \theta,
\ee where we have taken $\n=i n$ to obtain a Riemannian metric.
Switching off the Taub-NUT parameter yields the relevant static
Kottler potential, and in the limit $l\rightarrow \infty$ we obtain
the usual $\Lambda=0$ potential. Indeed, in this $\Lambda=0$ case,
the TN solution was demonstrated in \cite{GH} (see also section
\ref{sec:over}) to be of the Papapetrou class: given $\omega$ in
(\ref{GHoff}), it is possible to  show that the relevant Weyl
potential
--- $F(r)$ in (\ref{bob}) --- is also a function of $\omega$ (note
already that $\omega$ and $F$ only depend on $r$ unlike $A$).
Furthermore, this result ties in with the fact that, quite
generically, Papapetrou type solutions have non-trivial asymptotic
properties. Indeed, note that the $\theta$ dependent $A$ potential
in (\ref{TN4}) is non-vanishing in the large $r$ limit.

When $\Lambda\neq 0$, we can longer do this trick since the
Papapetrou ansatz works only for $\Omega=\Omega(A)$ and there is no
duality relation to take us to $\Omega=\Omega(\omega)$. This fact,
following the integrable cases of \cite{Charmousis:2003wm}, limits
the solutions to be in one of three classes (see the discussion at
the end of subsection \ref{subsec:papa}). For class I solutions in
particular, there is an extra Killing vector field and the solutions
in question are stationary and cylindrically symmetric. When the
extra Killing vector is null we  obtain $pp$-wave solutions
\cite{SKMHH}. Such solutions can be obtained directly combining the
results of the previous section with \cite{Charmousis:2003wm}. To
illustrate the method we restrict ourselves here to a simple
example. Start with the Weyl spacetime \cite{Charmousis:2003wm,ho}
\be \label{ruth} ds^2=(\cosh(k x))^2\left[-y^2 V dt^2 +{dy^2\over V
k^2y^2} +y^2
  dz^2\right]+ dx^2,
\ee where $k$ is the adS curvature, and the potential $V$ given by
\be V(y)=1-{M \over y^{2}}. \ee The solution is regular at the adS
horizon and there is an event horizon at $V=0$. It describes a
$3-$dimensional planar BTZ black hole embedded in a locally
$4-$dimensional adS spacetime. Furthermore the metric (\ref{ruth})
is a solution of the Weyl system (\ref{W1}-\ref{W3}) with \be
\alpha=(\cosh(kx))^2 y^2 \sqrt{V}, \qquad e^{2\varphi}=V,\qquad
e^{2\nu}=\alpha^{1/2}(\cosh(kx))^2. \ee According to
\cite{Charmousis:2003wm} it is a Class III solution since
  $\alpha$ is a function of $x$ and $y$.
It is now straightforward to calculate $A$ and $\Omega$ (\ref{om})
for the stationary version (\ref{melina}). Here, for simplicity, we
take $k_0=0$ obtaining \be A=\frac{k_1 V}{1-V}, \qquad
e^{\Omega}=\frac{k_1^2 V}{(1-V)^2}. \ee Thus metric (\ref{melina})
reads \be ds^2=\cosh^2(k x) \left( \frac{1}{(y^2-M)k^2}dy^2
-dt^2-\frac{2}{\sqrt{M}}(y^2-M)dt d\phi + (y^2-M)d\phi^2 \right)
+dx^2. \ee

It is also possible to construct a deformed adS soliton (or planar
black hole) as we will explicitely show for  $D=5$ in the next
section.

\section{Examples in $D=5$ dimensions}\label{sec:5}

As we stressed earlier, the EMD $d=4$ system and the uplifted $D=5$
system have unique properties: in particular, the duality relation
(\ref{duality}) applies even in the presence of a cosmological
constant, and it can be used to bring Ernst's equation into its
usual $\Lambda=0$ form. The duality can also be used in relation to
Papapetrou's method. Last but not least, for $\Lambda=0$ we can
construct an infinity of solutions seeded from given $D=4$
stationary and axisymmetric solutions. We examine these properties
one by one giving examples as we go along to illustrate them.

Our general starting point is again the electric EMD system
(\ref{WKK1}-\ref{WKK5}), which reads for $d=4$ and for arbitrary
couplings $\gamma$ and $\delta$: \bea
\Delta \alpha &=& -2 \Lambda \alpha^{\frac{1}{2}-\frac{\delta^2}{2}} e^{2\nu}, \\
\label{grandA}
\del \cdot \left (e^\Omega \alpha^{\gamma \delta} \del A \right ) &=& 0, \\
\label{Omega}
\frac{1}{\alpha} \del \cdot \left(\alpha \nabla \Omega \right ) + \frac{1+\gamma^2}{2\alpha^{1-\gamma\delta}} e^\Omega (\nabla A)^2 &=& 0, \\
\label{Psi}
\nabla \cdot \left ( \alpha \del \Psi_* \right ) &=& 0 , \\
 2 \nu_{,u} \frac{\alpha_u}{\alpha}-
\frac{\alpha_{,uu}}{\alpha} &=& \frac{1}{\gamma^2+1}
\left(2\Psi_{0,u}^2 +\frac{1}{2}  \Omega_{,u}^2 \right)+ \nonumber \\
&+& \frac{\epsilon}{2} e^{\Omega} \alpha^{\gamma\delta -1}  \left (
A_{\, ,u} \right )^2, \; (u\leftrightarrow v) \label{int} \eea A
solution is thus given by a set of functions ($\alpha$, A, $\Omega$,
$\Psi_*$), such that the dilatonic metric for arbitrary $\gamma$ and
$\delta$ reads \be \label{dila} ds^2=(d\r^2+d\z^2) \; e^{2\nu}
e^{2\delta\Psi_*\over{\gamma^2+1}} e^{{\gamma\delta \Omega} \over
 { \gamma^2+1}}\alpha^{{\delta^2-1}\over {2}}+ e^{{(\Omega-2\gamma\Psi_*
 )}\over {\gamma^2+1}}\alpha
 \left(e^{{2(2\gamma\Psi_*-\Omega)}\over{\gamma^2+1}}d\varphi^2+d\psi^2
 \right)
\ee with dilaton $\phi={{\gamma\Omega+\sqrt{2}\Psi_*}\over
{1+\gamma^2}}+\delta \ln \alpha$ and potential $A$. According to
(\ref{KKmetric}) and for specific values $\gamma = 1/\delta =
\sqrt{3}$, this corresponds to a $D=5$
 dimensional stationary spacetime
\be
\lb{kerrr} ds^2=(d\r^2+d\z^2) \; e^{2\nu}
\alpha^{-2/3}+\alpha^{2/3}\left\{e^{-{\Psi_*\over \sqrt{6}}}
[-e^{{\Omega\over 2}}(dt+A d\varphi)^2+ e^{{-\Omega\over 2}}
d\varphi^2 ]+ e^{{\sqrt{2}\Psi_*\over \sqrt{3}}}d\psi^2\right\}.
\ee

Let us dwell on the duality map (\ref{duality}). First, it is
 important to note the  $\gamma$ and $\delta$ dependence of the
 field equations when spacetime is stationary and $\Lambda\neq 0$. Since the
 duality takes us from a $\gamma \delta=1$ spacetime to $\gamma\delta=-1$
 spacetime it cannot be used to map between 5 dimensional solutions. A $D=5$ dimensional
 stationary and axisymmetric spacetime will be transformed into a $d=4$ static
 and axisymmetric solution with scalar and magnetic/electric charge.

Suppose, however, that we have instead a $D=5$ {\it static} spacetime
i.e.~$A=0$. This corresponds to some Weyl solution with
cosmological constant \cite{Charmousis:2003wm}. In that case the map
(\ref{duality}) indeed takes us from a $D=5$ dimensional to a $D=5$ solution. This is obvious from the
 form of the action (\ref{action}). A sign change of $\delta$ can always be
 compensated by a sign change of the scalar field $\phi$.
 For the metric, consider $\gamma
 = -1/\delta = \sqrt{3}$ whereupon the $D=5$ solution now reads
\be \lb{kerrrd} ds^2=(d\r^2+d\z^2) \; e^{2\nu}
\alpha^{-2/3}+\alpha^{2/3}\left\{ e^{-{\Psi_*\over \sqrt{6}}}\left(
-e^{{-\Omega\over 2}}dt^2+e^{{\Omega\over 2}} d\psi^2\right) +
e^{{\sqrt{2}\Psi_*\over \sqrt{3}}}d\varphi^2\right\}. \ee The
duality ${\A}$ which takes us back to the static version of
(\ref{kerrr}) is simply a double Wick rotation. As an example, the
$D=5$ adS
 Schwarzchild solution,
\be \label{schw} ds^2=r^2\left({dr^2\over r^2
V(r)}+d\theta^2\right)- V(r) dt^2 + r^2 \cos^2\theta d\varphi^2 +r^2
\sin^2 \theta d\psi^2 \ee with $V(r)=1-{\Lambda\over 3} r^2 -
{\mu\over r^2}$, is transformed by $\A$ into \be
ds^2=r^2\left({dr^2\over r^2V(r)}+d\theta^2\right)+ V(r) d\psi^2 -
r^2 \cos^2\theta dt^2 +r^2 \sin^2 \theta d\varphi^2 \ee which is
nothing but the adS soliton \cite{mers}.

A stationary rather than static example is the 5 dimensional
$\Lambda$-Kerr solution of Hawking et.~al \cite{hawking} with a single
angular momentum. The metric reads
\bea
ds_5^2&=&-\frac{\Delta}{\rho^2}\left(dt-\frac{a\sin^2\theta}{\Xi_a}d\varphi\right)^2
+\frac{\Delta_\theta\sin^2\theta}{\rho^2}\left(adt-\frac{r^2+a^2}{\Xi_a}d\varphi\right)^2\nonumber\\
&+& \rho^2 \left({dr^2\over \Delta
  }+\frac{d\theta^2}{\Delta_\theta}\right)+r^2 \cos^2\theta d\psi^2 \lb{KADS}
\eea
with $a$, $M$, $k$ the angular momentum parameter, the mass, and adS
curvature scale respectively and
\bea
\Delta&=&(r^2+a^2)(1+k^2 r^2)-2M,\\
\Delta_\theta&=&1-a^2k^2\cos^2\theta,\quad\Xi_a=1-a^2k^2,\\
\rho^2&=&r^2+a^2\cos^2\theta,\quad\Lambda=-6k^2.
\eea
Using (\ref{kerrr}) and applying the same trick as in (\ref{carter}) it is straightforward to identify the components,
\bea
\label{andrew}
\alpha &=& \frac{r\cos\theta\sin\theta}{\Xi_a  }
\sqrt{\Delta\Delta_\theta},\\
A&=& {a\sin^2 \theta (\Delta-\Delta_\theta(r^2+a^2))\over {\Xi_a
    (a^2\Delta_\theta \sin^2 \theta-\Delta)}},\nonumber\\
e^{\Omega}&=&\frac{\Xi_a^2 (\Delta - a^2\Delta_\theta \sin^2
  \theta)^2}{\Delta\Delta_\theta \rho^4 \sin^2 \theta } , \nonumber\\
\e^{2\nu}&=& \rho^2 \alpha^{2/3} , \nonumber\\
e^{-\frac{\sqrt{3}\Psi}{\sqrt{2}} }&=& \frac{\tan \theta}{\Xi_a  r^2 \cos \theta}  \sqrt{\Delta\Delta_\theta}, \nonumber
\eea
where implicitly we perform the coordinate transformation  $d\r=dr/\sqrt{\Delta}$ and $d\z =
d\theta/\sqrt{\Delta_\theta}$. In order to use the duality we need
to evaluate the dual potential (\ref{coffee}), $\omega$, defined by
$(-\partial_{\z}\omega,
\partial_{\r}\omega)=e^{\Omega} \alpha (\partial_{\r} A,
\partial_{\z} A)$ (recall that in 5D, ${\gamma\delta}=1$). We obtain
\be \label{olk} \omega=-\frac{a \cos^2\theta}{\rho^2}(\mu+k^2
r^2\rho^2), \ee The duality map (\ref{duality}) then takes us to a
$d=4$ EMD solution with $\gamma\delta =-1$: \be \label{isabelle}
ds^2=\frac{\sqrt{\Delta \Delta_\theta}\rho
  \sin\theta}{\sqrt{\Delta-a^2\Delta_\theta
  \sin^2\theta}}\left[\frac{\rho^2}{\Xi} \left({dr^2\over \Delta
  }+\frac{d\theta^2}{\Delta_\theta}\right) +\frac{\Delta-a^2\Delta_\theta
  \sin^2\theta}{\rho^2}d\psi^2+r^2 \cos^2\theta  d\varphi^2 \right]
\ee with scalar field
 \be \label{yo}
e^{\frac{2\phi}{\sqrt{3}}}=\frac{\Xi^2(\Delta-a^2\Delta_\theta
  \sin^2\theta)}{\Delta \Delta_\theta \rho^2 \sin^2\theta}.
\ee and potential (\ref{olk}). The (electric) Ernst potential
(\ref{Em1}) for the rotating black hole (\ref{KADS}) is \be
\E_-=\frac{r\cos\theta (\Delta - a^2\Delta_\theta \sin^2
\theta)}{\rho^2 }-i \frac{a \cos^2\theta}{\rho^2}(2M+k^2 r^2\rho^2)
\ee and shares a rather similar form to its 4 dimensional rotating
counterpart (\ref{brandon}).

As discussed in section 3.1, a
convenient way to generate solutions using the Ernst potential \cite{Ernst} is to set
\be
\label{psira}
\E_{+} = \frac{\xi -1}{\xi + 1} \, ,
\ee
where $\xi$ is a complex field depending on $(\r,\z)$.
In terms of $\xi$, equation (\ref{zernstM}) now reads
\be
\label{cplxernst} \frac{1}{\alpha} \del \cdot \left ( \alpha \del
\xi \right ) =
    \frac{2\xi^* \left ( \del \xi \right ) ^2}{|\xi|^2 -1} \, ,
\ee
where a star denotes complex conjugation. In this representation of the
potential, (\ref{cplxernst}) is invariant under the complex transformation (\ref{apc}).
Therefore a simple trick is to start with a real
Ernst potential say, $\E_+$, for $D=5$ in other words a Weyl solution. Let us
take an adS/Sch solution (\ref{schw}) as an example.
We have
\be
\E_+=e^{-\Omega/2}=\sqrt{V\over{r \sin \theta}}
\ee
and therefore, from (\ref{psira}),
\be
\xi=\frac{\sqrt{V}+r\sin \theta}{r\sin \theta -\sqrt{V}}.
\ee
Then we can apply
(\ref{apc}) for a convenient phase say $\vartheta=\pi/2$ in order to obtain an
    imaginary  $\xi$.
The newly generated Ernst potential from (\ref{psira}) is now complex and we find
\be
\label{kof}
(e^{-\Omega/2},A)=\left(\frac{2\sqrt{V}r \sin{\theta}}{V+r^2 \sin^2 \theta }, -\frac{V-r^2 \sin^2 \theta }{V+r^2 \sin^2\theta} \right).
\ee
Inserting this Ernst pair $(\Omega, A)$, together with $\alpha$, $\nu$ and $\Psi_*$
from (\ref{schw}), into (\ref{kerrr}) gives a rotating solution.

Let us now briefly present an example solution following the
Papapetrou method. Our starting point this time is a Class II
solution of \cite{Charmousis:2003wm} \be \label{weyl2d} ds^2=
e^{{2\over {3}}r}\left(- e^{-{2\sqrt{2}\over 3}\z}dt^2 +e^{{2\over
3\sqrt{2}}\z} (dx_{1}^2+dx_2^2) \right) +{1\over
-2\Lambda}(d\r^2+d\z^2) \ee which is static and axially symmetric.
Setting for simplicity $k_0=0$ and $l_0=0$ we obtain $\Psi=0$ and
from (\ref{tofv}) and (\ref{dav}) \be \alpha=e^\r, \quad \varphi=
\sqrt{2\over3}\z, \quad
A=\frac{k_1e^{\frac{{-2\sqrt{6}}\z}{3}}}{1-e^{\frac{-2\sqrt{6}}{3}z}}
\ee upon which using (\ref{melina1}) gives a solution in rotating
coordinates.

As we mentioned in Section 3.3, a special class of solutions can be
found by supposing that $\alpha$ and $\nu$ only depend on $\r$
whereas $\Omega$, $\Psi$ and $A$ only depend on $\z$. The
$\r$-dependent part is given by (\ref{coc}) and is the same as in
\cite{bowcock,tony} {\footnote{though there the fields
$\Omega=A=\Psi=0$ since the $D-2$ dimensional subspaces are of
maximal symmetry.}}. From (\ref{PD2}-\ref{PD5}), we deduce the
second subsystem for the $\z$ dependent part \ba \label{spemaxwell}
\left (e^\Omega \dot{A} \right )^{\cdot} &=& 0 \\
\label{speOmega}
\ddot{\Omega} + 2 e^\Omega \dot{A}^2 &=&0\\
\label{spepsi}
\ddot{\Psi} &=& 0\\
\label{speint} 2 \dot{\Psi}^2 +  \dot{\Omega}^2 &=& 4e^\Omega
\dot{A}^2 \, , \ea where a dot now stands for a derivative with
respect to $\z$. From (\ref{spepsi}), we deduce \be \label{psispe}
\Psi(\z)=\frac{\beta\z}{\sqrt{2}} \, , \ee where $\beta$ is some
real integration constant and we have taken $\Psi(0)=0$ as a choice
for the origin of the $\z$ coordinate. Now, from (\ref{spemaxwell})
\be \label{speA} \dot{A} = \lambda e^{-\Omega} \, , \ee where
$\lambda$ is a real integration constant. Substituting
(\ref{psispe}) and (\ref{speA}) into (\ref{speint}), we get \be
\label{Omegaspe} \half \dot{\Omega}^2 + \beta^2 =
4\lambda^2e^{-\Omega} \, . \ee When $\lambda=0$ and $\beta=0$ then
$A$, $\Omega$ are constant and $\Psi=0$. As we anticipated the
metric reduces to \be \label{yann} ds^2 = \left (
-\frac{\mu}{\r^2}+k^2 \r^2 \right ) d\z^2 +
\frac{d\r^2}{-\frac{\mu}{\r^2}+ k^2 \r^2 } + \r^2 \left (-dt^2 +
d\phi^2 + d\psi^2 \right ) \ee which is nothing but the planar adS
soliton{\footnote{By a suitable double Wick rotation one can get a
planar black hole with a compact Euclidean horizon.}} \cite{mers}.
When $\beta\neq0$ and $\lambda\neq0$ on the other hand we obtain a
non-trivial deformation of this solution. We get \be e^\Omega =
\frac{2\lambda^2}{\beta^2} \left (1 \pm \sin (\beta\z) \right
),\qquad A(\z) = \mp \frac{\beta}{2\lambda}  \, \frac{\cos
(\beta\z)}{1 \pm \sin(\beta\z)} \ee and the five dimensional metric
reads \ba ds^2 &=& - \r^2 e^{\frac{\beta\z}{2\sqrt3}} \sqrt{1\pm
\sin(\beta\z)} \left ( dt +
A(\z)  d\phi \right )^2 + \frac{d\r^2}{-\frac{\mu}{\r^2}+k^2 \r^2}  \nonumber \\
&& + \left (-\frac{\mu}{\r^2}+k^2 \r^2 \right ) d\z^2 + \r^2
e^{-\frac{\beta\z}{\sqrt{3}}} d\psi^2 + \frac{\beta^2}{2l^2} \r^2
\frac{e^{\frac{\beta\z}{2\sqrt3}}}{\sqrt{1\pm \sin(\beta\z)}}
d\phi^2 \ea We can go one step further by absorbing $\lambda/2\beta$
in $\phi$, renaming $\beta\rightarrow 2\beta$ and considering the
translation $\z \rightarrow \z+{\pi\over 2\beta}$. We get \bea
\label{ana} ds^2 &=& \frac{dr^2}{-\frac{\mu}{\r^2}+ k^2 \r^2}  +
\left (-\frac{\mu}{\r^2}+k^2 \r^2
\right ) d\z^2+ \nonumber  \\
     &+& \r^2 \left[
       e^{\frac{\beta\z}{\sqrt{3}}}|\cos(\beta\z)|\left[-dt^2+d\phi^2+2\tan(\beta\z)dt
d\phi \right]+ e^{-\frac{2\beta z}{\sqrt3}} d\psi^2 \right] \eea
This solution is clearly a continuous deformation of the adS soliton
which is obtained for $\beta=0$. The metric is not however
everywhere $C^2$; for every $z=n\pi+\frac{\pi}{2\beta}$ there is a
discontinuity in the first derivative with respect to $z$ which
indicates the presence of $\delta$ sources to account for these
jumps. The parameter $1/\beta$ indicates the distance between the
singularities. Also we can easily show that for $r=constant$ the
induced 4-dimensional metric is a vacuum solution to the 4
dimensional Einstein equations.  Surprisingly the deformed solution
(\ref{ana}) and the adS soliton have the same Krestschmann scalar
indicating that the deformed solution is again regular. Note that
the $\z$ coordinate varies throughout the real line because of the
exponential warp factors, which for $z$ negative and large
effectively reduce the $t-\phi$ dimensions, whereas for $z$ positive
and large, reduce the $\psi$ dimension. This solution has no 4
dimensional counterpart since the extra Weyl direction has to be
switched on (\ref{psispe}).

When $\beta=0$ but $\lambda\neq0$ we get \be A(\z) =
-\frac{1}{\lambda^2\z} \ee and the five dimensional metric reads \be
ds^2= \frac{d\r^2}{-\frac{\mu}{\r^2}+k^2 \r^2 } + \left (-
\frac{\mu}{\r^2}+k^2\r^2 \right ) d\z^2 + \r^2 \left (- \lambda\z
dt^2 + 2 dt\, d\phi +d\psi^2\right ). \ee Notice that, unlike the
previous case,  this solution can be Wick rotated to a non-static
black hole. Indeed, let us take \ba
\r &\rightarrow & i\r \\
\psi &\rightarrow  & i \psi \\
t &\rightarrow & \theta \\
\z &\rightarrow &  t \, , \ea to get \be ds^2 = - \left
(-\frac{\mu}{\r^2} +k^2 \r^2 \right ) dt^2 +
\frac{d\r^2}{-\frac{\mu}{\r^2}+ k^2 \r^2 } + \r^2 \left (  \lambda t
d\theta^2 + 2d\theta \, d\phi + d\psi^2 \right ). \ee Although this
metric has similar structure as the planar adS black hole the
horizon surface here has a non-trivial curved embedding depending on
the coordinate time
$t$. This solution is not a continuous deformation of the black hole
solution
and $\partial_\phi$ is a null Killing vector. This solution can also be
written in $D=4$ by simply taking $\psi=constant$ and using instead the 4
dimensional black hole potential.

\section{Constructing solutions for $\Lambda=0$}\label{sec:6}

As we pointed out in Section 2 we can construct solutions in a $D+n$
dimensional spacetime starting from a known seed solution in $D$
dimensions. This is possible as long as $\Lambda=0$. Say we start
from some 4 dimensional solution which can even be flat spacetime.
Then for each Weyl potential $\Psi$ solution of (\ref{PD4}) one can
construct a new $D=5$ dimensional solution finding the relevant
$\sigma$ component from (\ref{gif}). Schematically for each $D=4$
dimensional solution there is an infinity of $D=4+n$ dimensional
solutions that can be constructed, parametrised by the Weyl
potentials $\Psi_i$, $i=1,...,n$. A general analysis of this method
is best done in Weyl coordinates starting from lower to higher
dimension. Then, $\alpha=\r$ and we keep the same coordinate system
from lower to higher dimension. The Weyl potentials can be
constrained  in order to guarantee asymptotic flatness for the
higher dimensional solution. This we leave for later study. In this
section we will do the converse. Starting from  two 5 dimensional
examples, the Myers-Perry black hole and the black ring, we will go
down to 4 dimensions.

Start with the Myers-Perry solution describing a rotating black hole
with a single angular momentum in the coordinates (\ref{KADS}). We
set $\Lambda=0$, i.e. $k=0$, in the metric components (\ref{andrew})
and we use (\ref{basic}) to obtain \be \label{Expsigma}
e^{2\sigma}=\frac{(\rho^2-2M\sin^2 \theta)^{3/4}}{ \Delta^{1/12}
r^{2/3} \cos^{2/3} \theta \sin^{1/6}\theta}. \ee The corresponding
four-dimensional metric, solution of Einstein's equations, is given
by substituting the expressions for $\alpha$, $\Omega$, $A$ into
(\ref{andrew}) and (\ref{Expsigma}) in \be
ds^2=e^{2\sigma}\alpha^{1/6}\rho^2 \left(\frac{dr^2}{\Delta}
+d\theta^2\right)+\alpha e^{-\frac{\Omega}{2}}d\phi^2-\alpha
e^{\frac{\Omega}{2}} (dt+A d\phi)^2. \ee After the dust settles we
obtain the following 4-dimensional metric{\footnote{A word of
warning on notation. Here $\Delta=r^2+a^2-2M$
    stands for the 5-dimensional potential and thus the coordinate $r$ is
not
    the one appearing in (\ref{carter}).}}
\bea
&&ds^2 =\frac{\rho^2}{\sqrt{r\cos(\theta)}}[r^2+(a^2-2M)\sin^2
\theta]^{3/4}\left(d\theta^2+\frac{dr^2}{\Delta}
\right)-\nonumber\\
&&-
\frac{r\cos(\theta)(\rho^2-2M)}{\rho^2}\left(dt+\frac{2Ma\sin^2
\theta}{\rho^2-2M}
  d\phi\right)^2 +\nonumber \\
&&+\frac{\rho^2r\Delta \cos\theta\sin^2 \theta }{\rho^2-2M} d\phi^2
\nonumber. \eea Note that the resulting metric does not describe the
Kerr geometry and is not asymptotically flat (actually, even for
$M=a=0$ this solution is not flat).

Now we work out the seed solution for the black
ring solution \cite{ER} in $D=4$ dimensions. The black ring is described
in
C-metric type coordinates by the line element,
\bea
\label{blackring}
ds^2 &=& -\frac{F(y)}{F(x)}\left(dt+C(\nu,\lambda) \R
\frac{1+y}{F(y)}d\phi
\right)^2+ \nonumber \\
&+& \frac{\R^2
F(x)}{(x-y)^2}\left(-\frac{G(y)}{F(y)}d\phi^2-\frac{dy^2}{G(y)}+\frac{dx^2}{G(x)}+\frac{G(x)}{F(x)}d\psi^2
\right) \eea where $F(\xi)=1+\lambda \xi$,
$G(\xi)=(1-\xi^2)(1+\nu\xi)$, $\R$ is a constant giving roughly the
rings radius and \be \label{roberto}
C(\nu,\lambda)=\sqrt{(\lambda-\nu)\lambda\frac{1+\lambda}{1-\lambda}}.
\ee The radial and angular coordinates are respectively $y\in
]-\infty, -1]$ and $x\in [-1,1]$. A regular black ring without
conical singularity is obtained when the rotation cancels out the
gravitational attraction of the ring, for \be
\lambda=\frac{2\nu}{1+\nu^2} \ee In all other cases a conical
singularity naturally appears at $x=1$ holding the black ring
together and avoiding its collapse. Static black rings are obtained
when $\lambda=\nu$. This solution presents a lot of interesting
properties which are discussed in \cite{ER}, \cite{Harmark} and
\cite{rob2}.

The first thing we need to do is identify the components from
(\ref{kerrr}). We get,
\bea
\label{brcomp}
\alpha &=&\sqrt{-G(y)G(x)} \frac{\R ^2}{(x-y)^2}, \qquad
e^\Omega=\left(\frac{F(y)(x-y)}{F(x)\R}\right)^2 \frac{1}{-G(y)}\nonumber
\\
e^{\frac{\sqrt{3}\Psi_*}{\sqrt{2}}}  &=& \frac{\R
G(x)}{(x-y)\sqrt{-G(y)}}, \qquad A=C(\nu,\lambda) \R
\frac{(1+y)}{F(y)}\nonumber\\
e^{2\nu_{(5)}} &=& \alpha^{2/3}\frac{\R^2}{(x-y)^2}F(x) \eea To
construct the relevant $D=4$ solution from the above we keep the
same components $A$, $\Omega$ and $\alpha$ and we evaluate the
component $\sigma=\nu_{(4)}-\nu_{(5)}$ using (\ref{basic}) given the
components $\Psi$ and $\nu_{(5)}$ from (\ref{brcomp}). It is then
straightforward  to note that the $D=4$ metric (\ref{4d}) takes the
form, \be ds^2=e^{2\sigma}\alpha^{1/6}\frac{\R^2}{(x-y)^2}F(x)
\left(\frac{dx^2}{G(x)}-\frac{dy^2}{G(y)}\right)+\alpha
e^{-\frac{\Omega}{2}}d\phi^2-\alpha e^{\frac{\Omega}{2}} (dt+A
d\phi)^2 \ee and $\sigma$, in the above coordinate system, is given
by two first order ODE's (\ref{basic}) \bea \sigma_{,x} &=&
-\frac{\alpha}{8 (\alpha_{,x}^2 G(x)-\alpha_{,y}^2
G(y))}\left[\alpha_{,x}
  \{\Psi_{,x}^2 G(x)+\Psi_{,y}^2
  G(y)\}-2\Psi_{,x}\alpha_{,y}\Psi_{,y}
G(y)\right]\nonumber\\
\sigma_{,y} &=& \frac{\alpha}{8 (\alpha_{,x}^2 G(x)-\alpha_{,y}^2
G(y))}\left[\alpha_{,y}
  \{\Psi_{,x}^2 G(x)+\Psi_{,y}^2
  G(y)\}-2\Psi_{,x}\alpha_{,x}\Psi_{,y}
G(x)\right] \eea This can be integrated explicitly and we  obtain
\be e^{2\sigma}=\frac{(x-y)^{1/12} (W(x,y))^{3/4}}{(-G(y))^{1/12}
(G(x))^{1/3}} \ee where \be
W(x,y)=[y+x+\nu(1+xy)][\nu^2(xy-1)^2-[2+\nu(x+y)]^2]. \ee

\section{Conclusions}\label{sec:conc}

In this paper we have extensively analysed solution generating
methods for Einstein's equations in $D$ dimensions with a
cosmological constant. In particular, we studied stationary
spacetimes of axial symmetry, restricting our attention to the case
of a single rotation parameter. Our analysis was also shown to
apply, by a simple KK reduction, to an EMD
(Einstein-Maxwell-dilaton) system with a Liouville potential. Our
approach has been threefold.  Firstly, to make the connection with
the classical works of general relativity in $D=4$ and $\Lambda=0$
such as those of Papapetrou and Ernst, and also to connect with the
relatively few recent studies in higher dimensions for $\Lambda=0$
\cite{Harmark}. Our aim was to analyse the symmetries of the field
equations including possible dualities, to classify and characterise
the methods and solutions, and to give typical examples without
necessarily writing out all the possible metrics.

Our analysis of the field equations has brought out a new solution
generating method valid for $\Lambda=0$. According to this recipe,
for each 4 dimensional stationary and axisymmetric solution, one can
generate an infinity of higher dimensional solutions, parametrised
by a Weyl potential, for each extra dimension. In this way, even a
flat 4 dimensional solution can generate an infinite number of
higher dimensional solutions. As examples,  we showed that the 5
dimensional black ring and the 5 dimensional Myers-Perry solution do
not originate from Kerr's solution, the only stationary and
axisymmetric black hole solution in $D=4$. We have seen that this
method does not generically preserve asymptotic flatness. A more
systematic analysis of this method, in particular making use of the
Weyl coordinates (\ref{WC}), will be undertaken in the future. For
$\Lambda\neq 0$, we have found solutions which can be interpreted as
deformations of the adS soliton and planar black holes. These
solutions are of non-trivial topological charge characterised by an
extra integration parameter.

We have demonstrated that classical methods such as those of Ernst
and Papapetrou can be extended to spacetimes admitting a
cosmological constant. We generalised the results of Papapetrou
mapping a certain class of stationary solutions to static ones and
have found the extension of Ernst's equation in the presence of a
cosmological constant. We have seen that one can interpret the field
equations in a geometric way with respect to a three dimensional
background manifold. Whereas when $\Lambda=0$ the manifold in
question is flat, the presence of $\Lambda$ makes the manifold
curved and the choice of an adequate coordinate system difficult.
Our actual analysis leaves open the question of finding a suitable
coordinate system for asymptotically dS or adS spaces, such as those
available for $\Lambda=0$; namely that of spheroidal coordinates
\cite{Zipoy} or Weyl coordinates \cite{Weyl}. A coordinate system
adapted to the profile of the solution in question  would be able to
stretch the methods we have developed to their full potential. For
example, we would expect to be able to generate Carter's solution
\cite{carter} from Kottler's solution by a method similar to that
exposed by Ernst for the $\Lambda=0$ case \cite{Ernst}. The presence
of the cosmological constant has been shown here not to burden the
solution generating methods themselves, but rather to emphasise the
the adequate choice of a coordinate system with which to apply these
methods. This is of crucial importance in order to tackle solutions
such as the adS black ring, the black ring solution in higher
dimensions, or exact braneworld gravity solutions such as the black
hole on the brane (see for example \cite{kal}) or that of a cosmic
string \cite{davis}.

\section{Acknowledgements}

It is a pleasure to thank Roberto Emparan, Ruth Gregory and Bernard
Linet for discussions and especially Cedric Leygnac for
collaboration in the early stages of this work. CC and DL thank the
Galileo Galilei Institute for Theoretical Physics for their
hospitality and the INFN for partial financial support during the
completion of this work. CC also thanks Nemanja Kaloper during whose
visit part of this work was not completed.

\end{document}